\documentclass[a4paper,floatfix,pre]{revtex4-1}
\usepackage{hyperref}
\usepackage[british]{babel}
\usepackage[latin1]{inputenc}
\usepackage{amsmath,amssymb,mathtools}
\usepackage{bm}
\usepackage{graphicx}
\usepackage{tikz}
\usepackage{setspace}
\usepackage[sort&compress]{natbib}

\newcommand{\Sl}{^\circ\mathrm{S}}

\newcommand{\dd}{\mathrm{d}}
\newcommand{\Mov}{M_{ov}}
\newcommand{\Maz}{M_{az}}
\newcommand{\K}{\,\mathrm{K}}
\newcommand{\N}{\,\mathrm{N}}
\newcommand{\C}{^\circ\mathrm{C}}
\newcommand{\Sv}{\,\mathrm{Sv}}
\newcommand{\kg}{\,\mathrm{kg}}
\newcommand{\m}{\,\mathrm{m}}
\newcommand{\s}{\,\mathrm{s}}
\newcommand{\psu}{\,\mathrm{psu}}

\newcommand{\tot}[2]{\frac{\dd #1}{\dd #2}}
\newcommand{\EaMov}{E_a|_{\Mov}^0}
\newcommand{\EaqN}{E_a|_{q_N}^0}
\newcommand{\DqN}{D|_{q_N}^0}
\newcommand{\DMov}{D|_{\Mov}^0}
\newcommand{\qeqN}{q_{e}|_{q_N}^0}

\newcommand{\qNMov}{q_{N}|_{\Mov}^0}

\begin{document}

\title{
  Meridional overturning circulation: stability and ocean feedbacks in a box model.
}

\author{Andrea A. Cimatoribus}
\email{cimatori@knmi.nl}
\author{Sybren S. Drijfhout}
\affiliation{Royal Netherlands Meteorological Institute, De Bilt, The Netherlands}
\author{Henk A. Dijkstra}
\affiliation{Institute for Marine and Atmospheric research Utrecht, Utrecht University, Utrecht, The Netherlands}

\begin{abstract}

  A box model of the inter--hemispheric Atlantic meridional overturning circulation is developed, including a variable pycnocline depth for the tropical and subtropical regions.
  The circulation is forced by winds over a periodic channel in the south and by freshwater forcing at the surface.
  The model is aimed at investigating the ocean feedbacks related to perturbations in freshwater forcing from the atmosphere, and to changes in freshwater transport in the ocean.
  These feedbacks are closely connected with the stability properties of the meridional overturning circulation, in particular in response to freshwater perturbations.

  A separate box is used for representing the region north of the Antarctic circumpolar current in the Atlantic sector.
  The density difference between this region and the north of the basin is then used for scaling the downwelling in the north.
  These choices are essential for reproducing the sensitivity of the meridional overturning circulation observed in GCMs, and therefore suggest that the southernmost part of the Atlantic Ocean north of the Drake Passage is of fundamental importance for the stability of the meridional overturning circulation.
  With this configuration, the magnitude of the freshwater transport by the southern subtropical gyre strongly affects the response of the meridional overturning circulation to external forcing.

  The role of the freshwater transport by the overturning circulation ($\Mov$) as a stability indicator is discussed.
  It is investigated under which conditions its sign at the latitude of the southern tip of Africa can provide information on the existence of a second, permanently shut down, state of the overturning circulation in the box model.
  $\Mov$ will be an adequate indicator of the existence of multiple equilibria only if salt--advection feedback dominates over other processes in determining the response of the circulation to freshwater anomalies.
  $\Mov$ is a perfect indicator if feedbacks other than salt--advection are negligible.
\end{abstract}

\maketitle

\section{Introduction}
\label{sec:Introduction}

The stability of the Atlantic Meridional Overturning Circulation (MOC) in the ocean has been a topic of investigation in oceanography and climatology since the pioneering work of~\citet{stommel_thermohaline_1961}.
The central idea of \citet{stommel_thermohaline_1961} is that the MOC is linked to a feedback between the circulation and the salinity advected from the tropics to the subpolar regions, increasing density where sinking occurs and enhancing the MOC strength.
This mechanism, usually referred to as salt--advection feedback, is also responsible for the collapse and reversal of the circulation, if a sufficiently strong freshwater anomaly is dumped in the subpolar areas.
This concept was originally developed to describe the MOC in a single hemisphere, with the circulation being driven only by buoyancy forcing and, implicitly, by mixing--induced vertical diffusivity at low latitudes.
It has been used to describe the inter--hemispheric MOC as well, with the circulation being driven by a north--south, instead of equator--pole density difference~\cite{rooth_hydrology_1982,rahmstorf_freshwater_1996,huisman_indicator_2010}.
Even if extremely simple, this paradigm has been  helpful in understanding the behaviour of the MOC in more complex General Circulation Models (GCMs) of various complexity, which all show that the MOC can substantially weaken if a sufficiently strong perturbation is applied.

\begin{figure}[ht]
  \begin{center}
    \includegraphics[width=\textwidth]{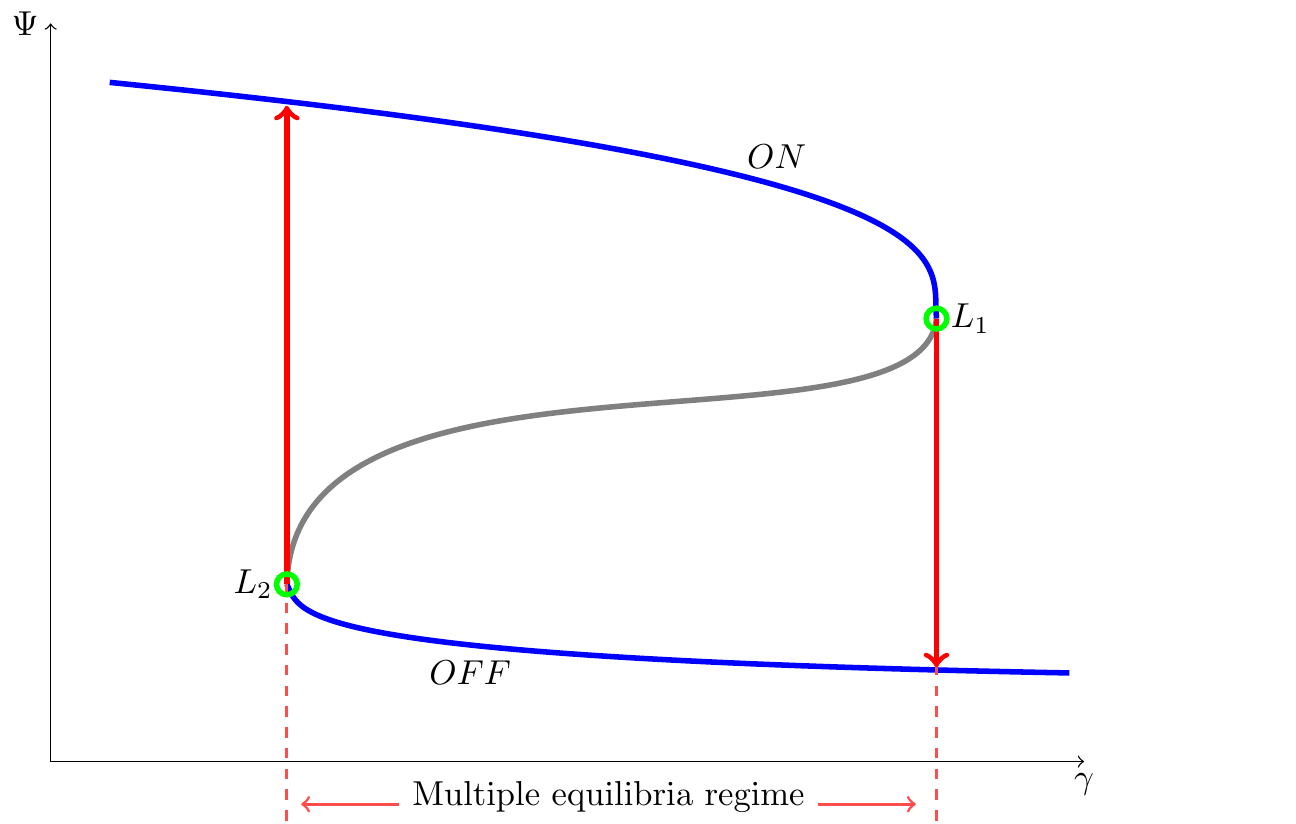} 
  \end{center}
  \caption{Sketch of a bifurcation diagram for the MOC strength $\Psi$, as a function of a generic freshwater forcing strength $\gamma$.
    The blue lines mark the two stable steady state solutions, ``ON'' and ``OFF''; the grey line marks the unstable solution connecting the two stable ones.
    At points $L_1$ and $L_2$, marked by green circles, the stability of the steady state solution changes.
    The range of $\gamma$--values between the two dashed vertical red lines is the ME regime, where two stable states coexist under the same boundary conditions.
    Changing $\gamma$ further, the system jumps from one solution to the other at the two limit points $L_1$ and $L_2$, as indicated by the red arrows.
    \label{fig:BifurcationSketch}}
\end{figure}

It has been shown in several earth system models of intermediate complexity that if the freshwater forcing over the North Atlantic (say measured by a parameter $\gamma$) is increased, the MOC will first slow--down and then undergo a more or less abrupt transition to a collapsed state~\cite{rahmstorf_thermohaline_2005}.
The point of collapse is believed to be associated, in the terminology of dynamical systems theory, to a saddle node bifurcation.
At such a point, which we will call $L_1$, a previously stable steady state of the system (the present day ``ON'' state of the MOC, see Fig.~\ref{fig:BifurcationSketch}) loses its stability, and the system jumps to another stable steady state with little or reversed overturning in the Atlantic (the ``OFF'' state, see Fig.~\ref{fig:BifurcationSketch}).
If the forcing is then reduced again, the MOC does not recover immediately, but stays collapsed until another critical value of the freshwater forcing is reached.
At this second point, named $L_2$, the ``OFF'' state loses its stability and the system jumps to the ``ON'' state, with vigorous overturning.
In the region of $\gamma$ between $L_1$ and $L_2$, called the multiple equilibria (ME) regime, two stable steady states coexist under the same boundary conditions, separated by an unstable steady state of the system, usually not observable in numerical models performing time integration.

In this context, it has been suggested that the freshwater transport by the MOC at the southern border of the Atlantic Ocean, usually shorthanded $M_{ov}^{30\Sl}$, may play a special role in determining MOC stability \cite{de_vries_atlantic_2005,huisman_indicator_2010}.
In analogy with the original salt--advection feedback of the Stommel model, the sign of the freshwater transport may determine  whether the MOC could be permanently collapsed or, in other words, if a second stable steady solution, with a very weak or reversed MOC, exists.

If $M_{ov}^{30\Sl}$ is negative, the MOC is importing salt into the basin.
A negative perturbation on the MOC would determine a negative salinity anomaly, and consequently a negative density anomaly, within the Atlantic basin.
If sufficiently strong, this density anomaly can be advected northward up to the sinking regions of  the northern North Atlantic, either by the gyres or by the MOC itself.
There, it would feed back onto the MOC strength, reducing the downwelling rate and amplifying the initial negative MOC perturbation, providing a mechanism for the collapse of the MOC.
If, on the other hand, $M_{ov}^{30\Sl}$ is positive, the MOC is importing freshwater into the Atlantic basin.
A negative perturbation in MOC strength would then lead to a positive density perturbation, as it would bring a positive salinity anomaly in the Atlantic, and no amplification of the initial perturbation is possible.
In this second case, even if the MOC can be slowed down substantially or even reversed by a sufficiently strong freshwater perturbation, it will spontaneously recover when the perturbation is removed, as no stable ``OFF'' state of the MOC exists under the unperturbed boundary conditions.

In \citet{de_vries_atlantic_2005}, it is shown that bringing $M_{ov}^{30\Sl}$ to negative values through appropriate surface freshwater flux anomalies it was possible to permanently collapse the MOC if net evaporation over the North Atlantic is below a critical value.
In particular, it was shown that a dipole freshwater anomaly applied over the southern portion of the Atlantic Ocean affects the freshwater transport by the southern subtropical gyre circulation (usually referred to as $M_{az}^{30\Sl}$), and hence can affect $M_{ov}^{30\Sl}$.
With an increase of $M_{az}^{30\Sl}$, $M_{ov}^{30\Sl}$ can eventually become negative, enabling a permanent shut--down of the MOC.

Recently, a series of numerical experiments with two different GCMs was performed to further explore the importance of freshwater budget of the Atlantic Ocean.
In \citet{cimatoribus_sensitivity_2011}, it is shown that the freshwater transport by the southern subtropical gyre is of paramount importance in determining the stability of the MOC, even if the overturning rate is only weakly sensitive to $M_{az}^{30\Sl}$.
Similarly to~\citet{de_vries_atlantic_2005}, $M_{az}^{30\Sl}$ and $M_{ov}^{30\Sl}$ were tuned through freshwater anomalies at the surface, showing that the sensitivity of the MOC to changes in the net precipitation over the Atlantic Ocean depends strongly on how the freshwater transport at the southern border of the Atlantic is divided between the gyre and the MOC.
In particular, as the freshwater transport by the gyre is increased, $M_{ov}^{30\Sl}$ decreases and this leads to a permanent collapse of the MOC for weaker reductions of the net evaporation.
The MOC can even be collapsed by solely increasing $M_{az}^{30\Sl}$, without changing the integrated net evaporation over the Atlantic basin.
A similar sensitivity of the MOC to the freshwater transport in the South Atlantic can be inferred from the work of~\citet{marsh_stability_2007}.
They show that, in a low resolution model, changes in the zonal salinity contrast over the South Atlantic control the MOC sensitivity to freshwater perturbations.

The results above consistently point to the importance of the freshwater (that is, buoyancy) forcing in setting the MOC stability and strength in connection with the salt--advection feedback, but they are in apparent contradiction with the point of view of~\citet{ferrari_ocean_2009}, who suggest that the direct mechanical forcing by the wind over the Southern Ocean is the main driver of the MOC.
A synthesis of these two views was attempted in~\citet{wolfe_adiabatic_2011} and \citet{nikurashin_theory_2012}; a series of numerical experiments with an ocean--only GCM (both at eddy--resolving and non eddy--resolving resolutions) contribute to a description of the MOC as driven by the wind--induced upwelling in the Southern Ocean, but with buoyancy distribution controlling whether an active pole--to--pole overturning can actually take place.
A similar synthesis of these two views was attempted in simple models by \citet{gnanadesikan_simple_1999} and \citet{johnson_reconciling_2007} (from now on J07).

On the other hand, the results of \citet{longworth_ocean_2005} point to the importance of the transport of salinity by the wind driven gyres (represented in a box model by horizontal diffusion) for the stability of the MOC.
In particular, they show that a simultaneous increase in the horizontal diffusivity in both hemispheres is in general a stabilising factor for the MOC, providing a way to transport salt from low to high latitudes that bypasses the salt--advection feedback mechanism.

In this paper, the main goal is to understand the results of \citet{cimatoribus_sensitivity_2011} in the framework of a box model, and to better understand the validity of $M_{ov}^{30\Sl}$ as an indicator of MOC stability.
The box model is based on that developed by J07 and \citet{gnanadesikan_simple_1999}, with a few important differences.
Its formulation and solution method are described in Sec.~\ref{sec:methods}.
We are particularly interested in understanding the role of the freshwater transport by the gyre and overturning circulation at the southern entrance of the Atlantic Ocean on the changes in stability of the MOC found in \citet{cimatoribus_sensitivity_2011}.
This can be done by including a representation of the transport by the southern subtropical gyre as a horizontal diffusive transport, similarly to what was done by~\citet{longworth_ocean_2005}.
The sensitivity of the solutions of the model to this as well as to other parameters of the system is described in Sec.~\ref{sec:Results}.
In relation to this, the conditions under which $M_{ov}^{30\Sl}$ can provide useful information on the stability properties of the MOC are discussed in Sec.~\ref{sec:Mov}.

\section{Methods}
\label{sec:methods}

\subsection{Model definition}

\begin{figure}[ht]
  \begin{center}
    \includegraphics[width=\textwidth]{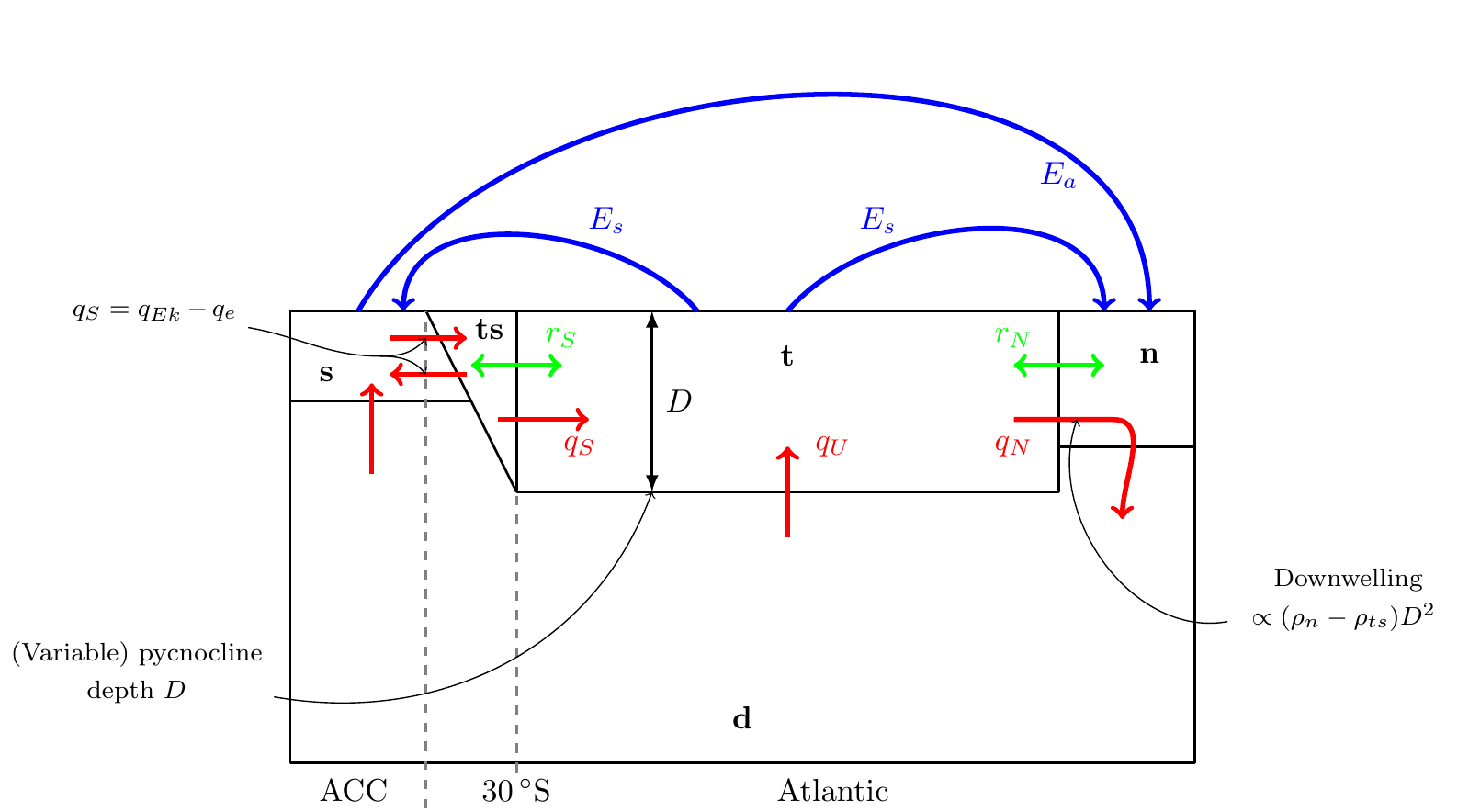}
  \end{center}
  \caption{Structure of the box model, with flow pathways connecting the different boxes.
    Red arrows represent net volume fluxes (with names in the same colour), green arrows represent gyre exchanges between the different boxes (with names in the same colour), blue arrows are water vapour transports through the atmosphere (with names in the same colour).
    In black, the boxes names.}
  \label{fig:Model}
\end{figure}

The box model (see Fig.~\ref{fig:Model}) used in this work consists of a basin spanning both hemispheres (which represents the Atlantic Ocean) connected in the south to a periodic channel (representing the Southern Ocean, shorthanded s).
It is assumed that the entire flow at depth from the Atlantic Ocean is upwelled into the mixed layer represented by box s, due to Ekman pumping.
Upwelling in other basins, due to diapycnal mixing, is thus neglected in this work.
The basin has four other boxes: a pycnocline (made of two boxes: t and ts) and a deep box (box d) separated by a pycnocline at a variable depth $D$, and a northern box (box n) of fixed volume.
The northern box only represents that portion of the Northern Seas where downwelling actually takes place.
Differently from J07 and \citet{gnanadesikan_simple_1999}, the Atlantic thermocline and the Southern Ocean are connected through a further box, named ts.
The box extends south of the latitude of the tip of the African continent (approximately $30\Sl$).
The use of this box, motivated in detail in the following sections, enables to compute the meridional density gradient within the Atlantic basin, assumed to be the driver of the overturning circulation (see Secs.~\ref{sec:qN} and~\ref{sec:boxts}).
This region is characterised by larger isopycnal slopes than within the Atlantic basin (see, e.~g.~\citet{lumpkin_global_2007}), with the pycnocline becoming shallower moving poleward.
The box ends in the south at the latitude where the deepest Antarctic Intermediate Waters (northward flowing) outcrop, approximately $40\Sl$~\cite{lumpkin_global_2007}.
The depth of this box at its northern end is the same depth scale $D$ used for the main thermocline box t.
This depth scale is obtained from the pycnocline model used by~J07 and~\citet{gnanadesikan_simple_1999}, but with the thermocline box divided in the two boxes t and ts, as described.
The temperature is fixed in all boxes, while the salinities are prognostic variables of the model.
This is a common choice in simple models of the MOC, stemming from the much shorter decay time of large scale temperature perturbations in the ocean, with respect to salinity ones.
This choice implies that we will neglect any feedback with the atmosphere.
This appears as an obvious limitation of the work; but it may not be, as recent studies suggest that atmospheric feedbacks play a secondary role for MOC stability~\cite{den_toom_effect_2012,arzel_impact_2010,scott_interhemispheric_1999}.

The pycnocline model results from a balance between the following terms:
\begin{enumerate}
\item Ekman transport into the Atlantic basin from the Southern Ocean (and through the ts box), associated with the wind--driven upwelling in the Southern Ocean ($q_{Ek}$);
\item eddy induced flow out of the Atlantic basin (from box ts to box s) due to baroclinic instability ($q_{e}$);
\item diffusively controlled upwelling of water from the deep layer ($q_U$);
\item downwelling in the North Atlantic ($q_N$).
\end{enumerate}
Within box t, we assume that the MOC can be approximated as a coherent flow, northward in the pycnocline and southward at depth.
This pycnocline model can be paralleled to the view of the MOC by~\citet{wolfe_adiabatic_2011}, which also includes a wind--driven upwelling (``pushed'' from the south) and a diffusively controlled upwelling inside the basin, the latter compensating only for a small part of the downwelling in the north.
The volume of the Atlantic pycnocline is the sum of the volumes of boxes t and ts, and the volume budget of the pycnocline thus reads:
   \begin{equation}
      \begin{split}
         \frac{\dd(V_t+V_{ts})}{\dd t}
         &= \frac{\tau  L_{xS}}{\rho_0 |f_S|} - A_{GM}\frac{L_{xA}}{L_y} D +\frac{\kappa  A}{D} -q_N\\
         &= q_{Ek} - q_{e} + q_U - q_{N}\\
         &= q_S + q_U - q_{N},\text{ with } q_S=q_{Ek}-q_e.
         \label{eq:Pycnocline}
      \end{split}
   \end{equation}
In the equation, $\tau$ is an average zonal wind stress amplitude over the periodic channel region, $L_{xS}$ is the zonal extent of the Southern Ocean, $L_{xA}$ is the zonal extent of the Atlantic Ocean at its southern end, $D$ is the pycnocline depth, $\rho_0$ is a reference density, $f_S$ is the Coriolis parameter in the frontal region of the Southern Ocean, $A_{GM}$ is an eddy diffusivity, $L_y$ is the meridional extent of the frontal region, $A$ is the horizontal area of the Atlantic pycnocline and $\kappa$ is the vertical diffusivity.
The total volume flux at the southern boundary of boxes t and ts is called $q_S$; it is the difference between the Ekman inflow and the outflow due to baroclinic instability, while $q_N$, the downwelling flux in the north, is left unspecified for the moment.
As the Ekman inflow in the Atlantic Ocean is related to the wind forcing over the whole ACC \cite{allison_where_2010,friocourt_water_2005}, the zonal width of the entire Southern Ocean is used in the computation of $q_{Ek}$.
The zonal width of the Atlantic Ocean is used instead for computing the eddy outflow into the periodic channel.
The volume of the box t is given by $V_t=A D$, while that of box ts by $V_{ts}=L_{xA} L_y D/2$.
It can be seen from Eq.~\eqref{eq:Pycnocline} that the volume flux from the box s goes through box ts into the main thermocline without any change, we thus neglect any diapycnal upwelling within box ts.
As far as the pycnocline model is concerned, boxes t and ts behave as a whole, and the model is in fact completely equivalent to that of J07 and \citet{gnanadesikan_simple_1999}.

The model can be interpreted as an analogue of the adiabatic pole--to--pole circulation paradigm of \citet{wolfe_adiabatic_2011} and \citet{nikurashin_theory_2012}.
Diabatic fluxes mainly take place in the north, at the interface between boxes t and n, where thermocline waters are made denser by the interaction with the atmosphere, and in the Southern Ocean, where water flowing northward in the Ekman layer gains buoyancy.
In the interior, the flow is mainly adiabatic, and the lower branch of the overturning is virtually isolated from the upper one, apart from the diffusive upwelling $q_U$, very small in our configuration.
The upwelling in the south is also adiabatic, as water mass transformation only takes place when the lower branch of the MOC enters the box s, the mixed layer of the ACC.

\subsubsection{The scaling for $q_N$}
\label{sec:qN}

Based on geostrophy, J07 used the scaling $q_N\propto D^2 (\rho_d-\rho_t)/\rho_0$ for the downwelling \cite{oliver_can_2005,johnson_theory_2002,gnanadesikan_simple_1999}, with $\rho_d$ the density of the deep box and $\rho_t$ that of the thermocline box.
Since at steady state the densities of the deep and northern boxes are equal, this scaling can be written as $q_N\propto D^2 (\rho_n-\rho_t)/\rho_0$, with $\rho_n$ the density of the northern box.
The density difference between the northern and pycnocline boxes is almost entirely determined, in the box model, by the temperature difference (fixed in J07 and here).
Using this scaling, the sensitivity to freshwater fluxes is thus too low (more than $1\Sv$ is needed to collapse the MOC---not shown).
Furthermore, if a representation of the southern subtropical gyre is included in the box model of J07, the MOC shows no sensitivity to the gyre strength and to the freshwater transport associated with it (not shown).
This latter result is in striking disagreement with the results of~\citet{cimatoribus_sensitivity_2011}, as discussed in section~\ref{sec:Introduction}.

\citet{levermann_atlantic_2010} tested the scalings used in J07 and~\citet{gnanadesikan_simple_1999}.
They show that the downwelling can be described by $q_N=\eta D^2 \Delta \rho$, with $\Delta \rho$ the density difference between north--western and tropical water at the base of the thermocline.
Even if the validity of the above results for different models is still uncertain \cite{de_boer_meridional_2010}, the connection between overturning strength and meridional pressure gradients is well established (see e.~g.~\citet{sijp_key_2012,cessi_eddy-driven_2009}).
In the simple context of the box model, this translates into a meridional density gradient in the thermocline~\cite{oliver_can_2005}.
We exploit this idea by taking $\Delta \rho=(\rho_n-\rho_{ts})/\rho_0$.
In our model, the box ts is the southern end of the thermocline, which means that the overturning circulation is proportional to the density difference between the sinking regions in the north and the southern end of the thermocline.
The pycnocline model then reads:
\begin{equation}
  (A + \frac{L_{xA} L_y}{2}) \tot{D}{t} = \frac{\kappa  A}{D} + \frac{\tau  L_{xS}}{\rho_0 |f_S|} - A_{GM}\frac{L_{xA}}{L_y} D - \eta \frac{(\rho_n-\rho_{ts})}{\rho_0} D^2.
  \label{eq:PycnoclineNew}
\end{equation}

From a heuristic point of view, this scaling for $q_N$ implies that, when a large meridional density gradient is present, the sensitivity of the MOC is mainly determined by changes in pycnocline depth (neglecting the variable pycnocline depth leads in fact to unrealistic sensitivity to wind stress, not shown), while a freshwater flux in the north leads to a collapse due to the reduction of the inter--hemispheric density difference.
We thus view the collapse due to a freshwater perturbation as ``buoyancy driven'', even though the sensitivity of the MOC in the ``ON'' state is dominated by ``wind driven'' dynamics (e.~g.~changes in the Ekman inflow).
Even though we can collapse the MOC exploiting a buoyancy flux, reducing the wind stress over the Southern Ocean below a critical value also brings $q_S$ to zero (see Sec.~\ref{sec:WStress}), leaving a purely diffusive intrahemispheric MOC ($q_N=q_U>0$) similarly to what is suggested by the results of~\citet{wolfe_adiabatic_2011}.

\subsubsection{The role of the box ts}
\label{sec:boxts}

The pycnocline model of Eq.~\eqref{eq:PycnoclineNew} is equivalent to the one defined in J07, the only difference being our thermocline box split into two boxes, in order to emphasise the importance of the meridional density gradient.

Considering the salinity transport, instead, we proceed in a different way from J07.
We argue that at the southern border of the thermocline (the southern border of box ts in our model) the Ekman inflow and eddy outflow, whose difference gives the net volume flux, have to be taken into account separately concerning the salt transport.
The salt flux from box s to box ts has then to be written as $q_{Ek} S_s - q_e S_{ts}$.
In other words, even if the net volume flux and net density flux may sum to zero, there can still be exchange of salt between the Southern Ocean and the pycnocline.
In case of no net volume flux ($q_{Ek}=q_e$) the salt transport must be compensated in density by heat transport (not considered explicitly in this work).
This salinity transport must then be associated to transport of spiciness: it represents in this case an isopycnal salinity flux.
In general, for $q_{Ek}\ne q_e$, the net volume flux will be associated with density transport on top of this isopycnal salinity transport.
Although the Ekman and eddy fluxes may compensate globally, they will not do so locally, so that transport of properties is still present.
From box ts to box t, instead, we assume that the flow is coherent because the flow is already within the pycnocline layer, in geostrophic balance and with less prominent baroclinic eddy flow.
The transport of salt from box ts to box t can then be written simply as $q_S S_{ts}$ (for the case of a northward net flow).
Upwelling in the box s is also taken proportional to the net flow $q_S$.

A last connection between boxes ts and t is added through a diffusive constant, to represent in the simplest possible way the transport by the southern subtropical gyre, similarly to what was done by \citet{longworth_ocean_2005} and J07 (the latter only considered the northern subpolar gyre).
We argue that, if the MOC is sensitive to the difference between the sinking regions and the thermocline waters north of the ACC, the southern subtropical gyre may play a very important role in setting this density difference.
The argument in J07 of the effect of $r_S$ being negligible, at least in comparison with the gyre in the northern part of the basin, does not seem to be applicable in this context.

With this configuration, we neglect the interaction with the atmosphere of the area represented by box ts.
This choice is done for simplicity, but seems to be at least indirectly supported by \citet{garzoli_south_2011}, who show that little mass water transformation takes place for lighter water classes in this region (their Fig.~6).

A more important assumption in this work is the neglect of the exchanges between the Atlantic Ocean and the other Ocean basins, in particular with the Indian Ocean through Agulhas Leakage.
This is still a highly debated subject, outside the scope of the present paper.

\subsubsection{Salt conservation equations}

The equations needed to close the system are given by salt and volume conservation in each box.
They comprise the following set of equations for a state of the MOC similar to the present day one ($q_N>q_S>q_U>0$):
\begin{subequations}
  \begin{align}
    \tot{\left(V_t S_t\right)}{t} &= q_S S_{ts} + q_U S_d - q_N S_t + r_S (S_{ts}-S_t) + r_N (S_n-S_t) + 2 E_s S_0,\label{eq:box-modela}\\
    \tot{\left(V_{ts} S_{ts}\right)}{t} &= q_{Ek} S_s - q_e S_{ts} -q_S S_{ts} + r_S (S_t-S_{ts}),\label{eq:box-modelb}\\
    V_n \tot{S_n}{t} &= q_N (S_t-S_n)+r_N (S_t-S_n)-(E_s+E_a) S_0,\label{eq:box-modelc}\\
    \tot{\left(V_d S_d\right)}{t} &= q_N S_n - q_U S_d - q_S S_d, \label{eq:box-modeld}\\
    V_s \tot{S_s}{t} &= q_S S_d + q_e S_{ts} - q_{Ek} S_s - (E_s-E_a) S_0,\label{eq:box-modele}\\
    &S_0 V_{tot}-V_n S_n -V_d S_d -V_t S_t -V_{ts} S_{ts}-V_s S_s=0,\label{eq:box-modelf}
  \end{align}\label{eq:box-model}
\end{subequations}
with $V$ being the volume of the various boxes and $V_{tot}=V_n+V_t+V_{ts}+V_d+V_s$.
The last equation states salt conservation in the model.
$S$ is the salinity in the various boxes, $r_S$ is the gyre exchange between boxes t and ts, while $r_N$ is the gyre exchange between the pycnocline and northern box.
The salt transport by the gyres takes in this simple framework the form of a diffusive constant.
$E_s$ is the symmetric part of the atmospheric freshwater flux, from the pycnocline to the boxes n and s.
$E_a$ is the asymmetric part of the atmospheric freshwater flux, transferring freshwater from the Southern Ocean into the box n.
Physically, the asymmetry in the freshwater forcing is due to the different sizes of the boxes in the north and in the south, differences in precipitation amount connected with the fact that box n is closer to the pole than box s and differences in the runoff from the continents.
We will change $E_a$, changing the North--South density gradient, for collapsing the MOC.
In reality, possible sources of the freshwater flux are ice--sheets at high latitudes or Arctic sea--ice, not represented in the box model; these will generally cause a net change in the total freshwater content of the ocean.
Using $E_a$ as a ``hosing'' flux, enables to maintain a closed salinity budget, and thus to explore steady states of the system; the approach is similar to that used in GCMs, where ``hosing'' fluxes are usually compensated over the rest of the Ocean surface.
Both $E_s$ and $E_a$ are represented through a virtual salt flux.
Despite the different definition used here, the atmospheric freshwater fluxes $E_a$ and $E_s$ can be cast in a form equivalent to that used in \citet{longworth_ocean_2005} and \citet{scott_interhemispheric_1999}.
$S_0$ is a reference salinity and a linear equation of state is used, i.~e.~$\rho=\rho_0(1-\alpha(T-T_0)+\beta(S-S_0))$, where $\alpha$ and $\beta$ are the constant thermal expansion and haline contraction coefficients respectively, $T_0$ is a reference temperature.

To summarise, there are six differences in our model definition from J07: (i) a different choice of the box sizes, to account for our interpretation of the model as the Atlantic and Southern Oceans alone; (ii) the pycnocline layer is split in two meridionally at the latitude of the southern tip of Africa; (iii) the two components of salt advection between the southern box and the thermocline, due to the Ekman and eddy fluxes, are treated separately; (iv) the meridional density gradient used in the scaling of $q_N$; (v) the use of a term representing gyre exchange ($r_S$) between boxes t and ts and (vi) a slightly different parameter choice including lower vertical and eddy diffusivities.

\subsubsection{Reference solution}
\label{sec:StandardSol}

At steady state, if $q_N>q_S>0$, it follows that $S_d=S_n$ from Eq.~\eqref{eq:box-modeld}, as $q_S=q_N-q_U$.
The reference parameter values are shown in Tab.~\ref{tab:params}.
A solution for the system of Eqs.~\eqref{eq:PycnoclineNew} and~\eqref{eq:box-model} can be found using Mathematica software~\cite{wolfram_research_inc._mathematica_2010}, but only after substituting the parameters with their numerical values.
The solution for the reference configuration of Tab.~\ref{tab:params} is reported in Tab.~\ref{tab:sol}.

Similarly to J07, the reference solution consists of a flow northward in the boxes ts and t and southward at depth, downwelling from box n to box d and upwelling from box d to box s.
Diffusive upwelling takes place from box d to box t.
The northern box is saltier than the southern one under symmetric forcing, due to the asymmetry of the circulation itself (in other words, the circulation is maintained by the salt--advection feedback).
Differently from J07, the value of pycnocline depth is close to what commonly observed in the real ocean and in GCMs.
This is mainly due to smaller downwelling flux than in J07, closer to the present day estimate of $18.7\pm2.1\Sv$ \cite{cunningham_temporal_2007}, given by the different scaling for $q_N$ and by the smaller vertical diffusion considered here.
The value of the hydraulic constant $\eta$ used is the one estimated in \cite{levermann_atlantic_2010}, but we stress that this quantity is badly constrained since it will strongly depend on the choice of the temperature difference between boxes n and ts (or, in a more realistic framework on the locations used for the computation of the meridional density difference).
For this reason, we will study the model at different values of the hydraulic constant $\eta$ (0.5, 1 and 1.5 times the value given by \citet{levermann_atlantic_2010}), in order to assess the robustness of the results.

The temperature chosen for the box $T_{ts}$ is $10\C$, giving a difference of $5\C$ between the northern box and the southern end of the pycnocline.
This temperature is representative of the waters north of the periodic channel in the South Atlantic.
This temperature, or the temperature difference between boxes n and ts \footnote{Since the equation of state is linear, only temperature (or salinity) differences are physically relevant.}, is particularly important since it determines the strength of the MOC for fixed $\eta$ and a given salinity distribution.
The choice of this temperature difference was made in order to use the estimate of $\eta$ given by \citet{levermann_atlantic_2010}, the most recent we are aware of, and to obtain at the same time a realistic MOC rate.

A negative value of $E_a$ is used for the reference solution, to account for the different interpretation of the boxes in the model, with respect to J07.
A negative $E_a$ is needed in order to reduce the export of freshwater from the Atlantic Ocean, and also to reduce the net precipitation over the sinking regions of the North Atlantic, also considering the asymmetry of the sizes of box n and s.

\subsubsection{Freshwater budget}

In order to compare our results with those from more complicated models, we compute the equivalent freshwater budget of the ``Atlantic'' basin north of $30\Sl$ (see Fig.~\ref{fig:Model}).
With the definitions:
\begin{equation*}
  \begin{split}
    \Mov &= -\frac{1}{S_0} q_S \left(S_{ts} - S_d\right), \\
    \Maz &= -\frac{1}{S_0} r_S (S_{ts} - S_t),
  \end{split}
\end{equation*}
valid under the same conditions considered in Eqs.~\eqref{eq:box-model}, we find that the net steady evaporation from the Atlantic basin must be balanced by the freshwater import by the ocean circulation:
\begin{equation}
  (E_s - E_a) = \Mov + \Maz.
  \label{eq:MovMaz}
\end{equation}
This budget includes the dominant terms in GCMs of various complexity~\cite{cimatoribus_sensitivity_2011,drijfhout_stability_2010}.
Eq.~\eqref{eq:MovMaz} states that the freshwater transported through the atmosphere out of the basin must be balanced by the freshwater transport by the ocean circulation, split into its overturning ($\Mov$) and gyre ($\Maz$) components.
$\Maz$ can be controlled by varying $r_S$, and $\Mov$ will change in the opposite way, as long as the net evaporation is kept constant.
This behaviour, which is trivial in this model, has been observed in GCMs as well.
In~\citet{cimatoribus_sensitivity_2011}, $\Maz$ is changed using anomalies in the surface freshwater flux instead of perturbing the gyre strength, and $\Mov$ reacts in a similar way as here, compensating for the changes in the azonal transport and keeping the total budget closed.

\subsection{Solution method}
\label{sec:solution}

The system of equations~\eqref{eq:PycnoclineNew} and~\eqref{eq:box-model} is studied using the software AUTO-07p, which enables the continuation and bifurcation of solutions of systems of Ordinary Differential Equations (ODEs)~\cite{doedel_auto-07p:_2009}.
Only exact (within numerical accuracy) steady states are obtained with this technique, and no transient behaviour is studied in any part of this work.

The system of equations~\eqref{eq:PycnoclineNew} and~\eqref{eq:box-model} can be written as the autonomous ODEs system:
\begin{equation}
  \frac{\mathrm{d}\mathbf{x}(t,\mathbf{p})}{\mathrm{d t}}=\mathbf{G}(\mathbf{x}(t),\mathbf{p}),
  \label{eq:ODE}
\end{equation}
with $\mathbf{x}$ the $6$--dimensional state vector, $\mathbf{p}$ the vector containing the system parameters, and $\mathbf{G}$ a nonlinear mapping.
Given a steady state solution $\mathbf{x}_0$ of \eqref{eq:ODE} for a particular parameter set $\mathbf{p}_0$, AUTO-07p can track the evolution of the steady--state solution as the value of one parameter of the system is changed; the solutions can be stable or unstable, and their stability is determined as well.
The initial solution $\mathbf{x}_0$ can be analytical or numerical, as long as it is known with sufficient numerical precision.
The software is based on Newton's method for continuation of steady states; for further details, see~\citet{doedel_auto-07p:_2009} and references therein.
Special points, such as saddle node bifurcations, are detected and can be continued in two parameters, obtaining the locus of those special points in a parameter plane.
Multiple steady states can be identified following the solution beyond a special point and, more in general, a systematic exploration of the parameter space is possible.
All the results are obtained with small continuation step sizes, so that the results are not sensitive to the value actually used.
The relative convergence criterion for the steady solution in AUTO-07p is set to $10^{-7}$.

In the next section, starting from the reference solution obtained from Mathematica (see Sec.~\ref{sec:StandardSol}), the sensitivity of the model to several parameters is studied.
Solutions are continued within physically plausible regions of the parameter space, or down to the point where $q_S=0$.
When the sensitivity to freshwater fluxes is studied, the solution is continued for $q_S<0$ down to the point at which the downwelling in the north stops.
When $q_S$ changes sign, the definition of the box model has to be changed.
In particular, Eqs.~\eqref{eq:box-modela}, \eqref{eq:box-modelb}, \eqref{eq:box-modeld} and \eqref{eq:box-modele} become for a negative $q_S$:
\begin{equation*}
  \begin{aligned}
    \tot{\left(V_t S_t\right)}{t} &= q_S S_t + q_U S_d - q_N S_t + r_S (S_{ts}-S_t) + r_N (S_n-S_t) + 2 E_s S_0,\\
    \tot{\left(V_{ts} S_{ts}\right)}{t} &= q_{Ek} S_s - q_e S_{ts} -q_S S_t + r_S (S_t-S_{ts}),\\
    \tot{\left(V_d S_d\right)}{t} &= q_N S_n - q_U S_d - q_S S_s,\\
    V_s \tot{S_s}{t} &= q_S S_s + q_e S_{ts} - q_{Ek} S_s - (E_s-E_a) S_0,\\
  \end{aligned}
\end{equation*}
while the equations for the northern box and total salt conservation are unchanged.
The two sets of equations are automatically chosen in order to match the solution at $q_S=0$.

The  solutions  are not continued below $q_N=0$ as then the MOC is completely reversed and the scaling for $q_N$ is not meaningful anymore.
In fact, if $q_N$ was allowed to reach negative values, the scaling used in the ``ON'' state would represent an enhanced upwelling in the high latitudes of the North Atlantic, for which no plausible physical mechanism is known.
Furthermore, a reversed MOC is determined by the shallow outflow from the Atlantic, in the present box model this is due to an eddy flux stronger than the Ekman transport.
Whether such a situation could exist in reality is unclear.
With these limitations in mind, the study stops at the point where $q_N=0$ with $q_S$ small and negative.
The change in sign of $q_S$ when the downwelling in the north is weaker than the mixing--driven upwelling ($q_N<q_U$) does not affect the fundamental properties of the model, which is still controlled by the same scaling law for $q_N$.
On the contrary, the model is no more valid when $\Delta\rho$ and $q_N$ change sign.
An extension of the model including a consistent representation of the completely reversed MOC is beyond the scope of this work.

\section{Results}
\label{sec:Results}

This section is comprised of three subsections.
In subsection~\ref{sec:sensitivity} we investigate the sensitivity of the model solutions to the southern ocean wind stress and the strength of the southern gyre.
In subsection~\ref{sec:stability}, the stability of the MOC to freshwater perturbations is considered while in subsection~\ref{sec:Mov}, the results in~\ref{sec:stability} are interpreted using the freshwater budget of the box model. 

\subsection{Model sensitivity}
\label{sec:sensitivity}

In general, the sensitivity to the model parameters is very similar to that found in J07.
We will briefly discuss the importance of wind stress, and then analyse the sensitivity of the model to the gyre transport and to changes in the freshwater fluxes.

\subsubsection{Southern Ocean wind stress}
\label{sec:WStress}

\begin{figure}[th]
  \begin{center}
    \includegraphics[width=0.7\textwidth]{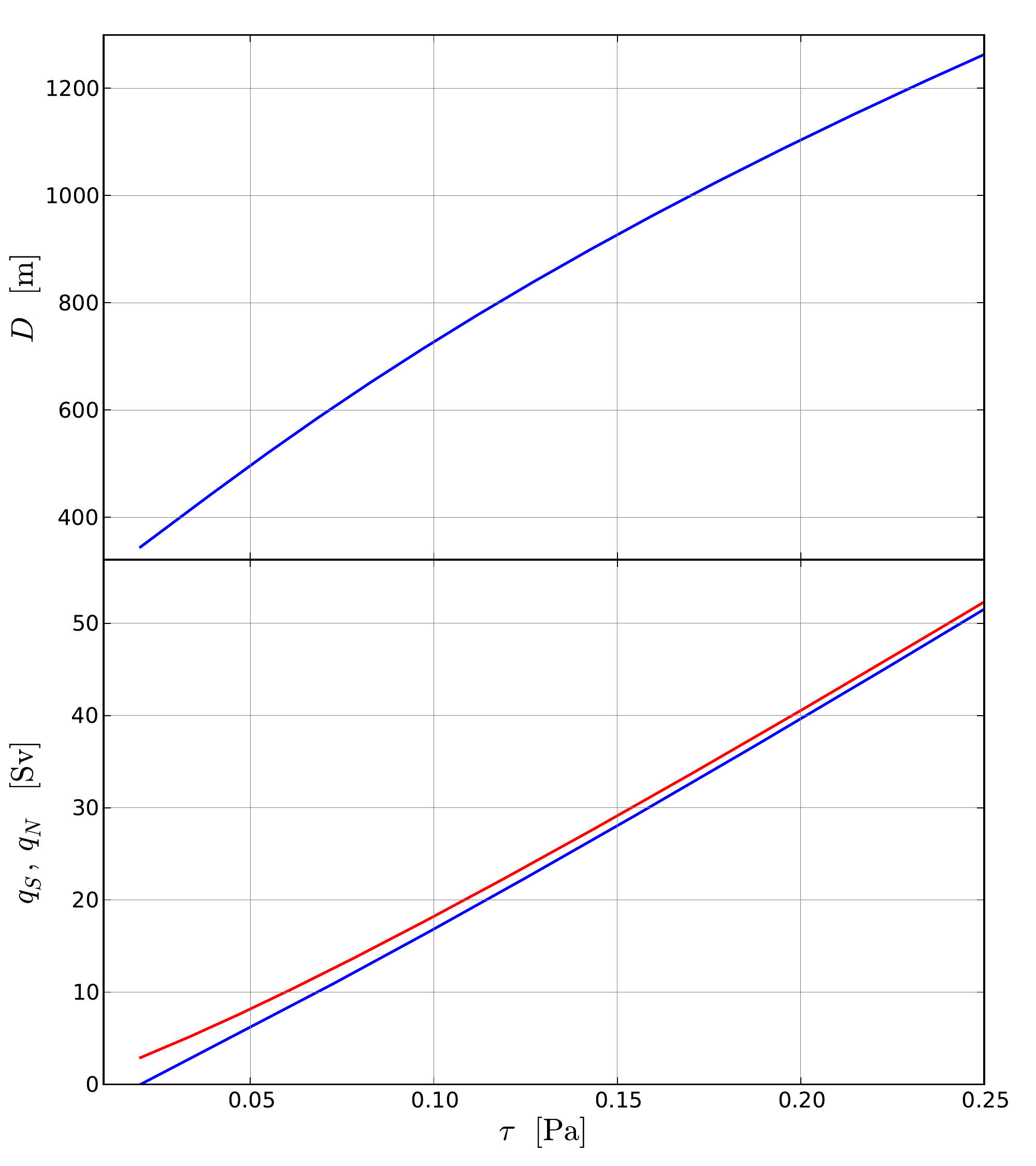}
  \end{center}
  \caption{Sensitivity of $D$ (top panel, blue), $q_S$ (lower panel, blue) and $q_N$ (lower panel, red) to the wind stress $\tau$.
    All other parameters are kept at the reference values of Tab.~\ref{tab:params}.}
  \label{fig:DqSqN_tau}
\end{figure}

In Fig.~\ref{fig:DqSqN_tau} the sensitivity of $D$, $q_S$ and $q_N$ to the wind stress $\tau$ is shown.
When the wind stress is changed, the role of pycnocline depth in determining the MOC strength is dominant.
The values of the volume transports change mostly due to the increased Ekman inflow at the southern border, generating a deeper pycnocline and a consequently stronger outflow in the north, which is qualitatively the same behaviour as in J07.
The changes in the meridional density difference, with a reduction of the salinity in the box ts (due to the increased inflow from the fresher box s) but little change in the northern box, enhance the changes in $q_N$ and $q_S$.
Below a certain critical value of $\tau$ (approximately $0.02\N \m^{-2}$) the eddy flux becomes as large as the Ekman inflow, and the waters sinking in the north are upwelled only through diffusion within the Atlantic basin.
Without a net inflow from the Southern Ocean, the MOC can not extend beyond the Atlantic basin, consistently with what is shown in the numerical experiments of~\cite{wolfe_adiabatic_2011}; no pole--to--pole MOC is possible without wind--stress over the Southern Ocean.

\subsubsection{Southern gyre}
\label{sec:rS}

As discussed in Sec.~\ref{sec:StandardSol}, a basic representation of the southern subtropical gyre is included in this work.
The sensitivity of $D$, $q_S$ and $q_N$ to $r_S$ is shown in Fig.~\ref{fig:DqSqN_rS}.
In addition, the sensitivity of $M_{ov}$ and $M_{az}$ is also shown in the bottom panel.
Both $q_S$ and $q_N$ decrease when the gyre strength increases, with stronger sensitivity for lower values of $r_S$.
This is a purely buoyancy driven response, which we could not reproduce using the J07 model with a term equivalent to $r_S$ added to their equations (not shown).
Downwelling in the north decreases due to the lowering of the north--south density difference, caused by the increased exchange between boxes ts and t which leads to a salinification of the southern end of the thermocline and a slight freshening of the box t.
The pycnocline depth increases moderately, in connection with the slightly stronger decrease of $q_N$ compared to $q_S$, and a slightly decreased upwelling flux $q_U$.
The northward volume transport into box ts is reduced by the stronger eddy outflow, associated with a deeper pycnocline.
This response is absent from J07 since $r_S$ strongly increases the salinity of box ts, but very little that of boxes t, n and d.
The scaling used in this work for $q_N$ is thus affecting the MOC (as it depends on $(\rho_n-\rho_{ts})$), contrary to the one used in J07 ($\propto (\rho_d-\rho_t)$).

As expected from the freshwater budget, Eq.~\eqref{eq:MovMaz}, $M_{ov}$ decreases and becomes negative as $M_{az}$ increases from zero, similarly to what is observed in numerical experiments with GCMs~\cite{huisman_indicator_2010,cimatoribus_sensitivity_2011,de_vries_atlantic_2005}.
This behaviour, trivial in the box model, has been exploited in GCMs to change $M_{ov}$, in order to reach the ME regime~\cite{de_vries_atlantic_2005,cimatoribus_sensitivity_2011}.
In GCMs, this behaviour is usually induced by different means; $M_{az}$ is increased not by modifying the gyre strength, but rather by perturbing the surface salinity, and $M_{ov}$ compensates $M_{az}$ due to changes in the intermediate depth ocean stratification~\cite{cimatoribus_sensitivity_2011}.

The response of $\Maz$ to $r_S$ saturates for large values of $r_S$.
This is due to the decrease in the salinity difference between boxes ts and t as $r_S$ increases, which limits the ability to tune $\Maz$ using $r_S$ in the box model.
Another limit of representing the gyre through a diffusive constant, already discussed by \citet{longworth_ocean_2005}, is the fact that the freshwater exchange depends on the meridional density difference while in reality the freshwater transport by the wind--driven gyre will rather scale with the zonal salinity gradient.

\subsection{Stability to freshwater perturbations}
\label{sec:stability}

\subsubsection{Symmetric freshwater flux}
\label{sec:Es}

Opposite to what found in J07, the MOC strength increases when $E_s$ is increased (Fig.~\ref{fig:DqSqN_Es}, the results are shown for three different values of $\eta$).
The different sign of the sensitivity to $E_s$ is due to the new definition of the $q_N$ scaling.
Increasing $E_s$, freshwater is moved from the thermocline box to the two high--latitude boxes.
Given the sense of the overturning circulation, this produces an increase in the density difference between boxes n and ts, as the freshening by the freshwater forcing in the box n is partly compensated by increased salt advection from the thermocline, while box ts is made fresher by its exchange with the southern box.
The response is thus due to the asymmetry between north and south, induced by the MOC itself; the sense of the circulation (northward above the pycnocline) causes a stronger salt transport from the thermocline towards the north than towards the south.
It is interesting to note that an increase of $E_s$ amounts to a decrease of the freshwater forcing over the entire Atlantic basin, as the net evaporation from the Atlantic (boxes ts, t and n) is given $(E_s-E_a)$.
The pycnocline depth decreases in response to the stronger outflow from box t to box n.

Extending the study to the non physical regime $E_s<0$ (net evaporation in the high latitudes) confirms the role of the MOC asymmetry; the MOC can not be collapsed decreasing $E_s$ below some threshold because, as the MOC weakens, also the impact of $E_s$ on the north--south density difference decreases.
The sensitivity of the MOC thus goes to zero for large enough negative values of $E_s$ (not shown).
Similarly, as the value of $r_S$ increases, the sensitivity of $q_N$ to $E_S$ decreases (not shown) as the southern subtropical gyre transports salty waters from box t to box ts regardless of the sense of the overturning circulation.

\subsubsection{Asymmetric freshwater flux}
\label{sec:Ea}

In Fig.~\ref{fig:DqSqN_Ea}, three bifurcation diagrams computed for different values of $\eta$ are shown, including $D$, $q_N$, $q_S$ and $M_{ov}$.

As expected, we find that an increase in the freshwater flux $E_a$ leads to a decrease of the MOC strength.
We thus recover the result of~\cite{den_toom_effect_2012,cimatoribus_sensitivity_2011}, that the MOC strength is controlled by the net evaporation of the overall Atlantic basin.
Even if the area where a freshwater anomaly is applied determines quantitative changes in the sensitivity of the MOC, the sign of this sensitivity is the same as long as the perturbation is applied within the basin.
In the framework of our box model, both increasing $E_a$ or decreasing $E_s$ amounts to a reduction of net evaporation out of the Atlantic basin, and both determine a decrease of the MOC strength.
An increase of $E_a$ is, however, much more effective in reducing the MOC strength than a decrease of $E_s$, since it changes directly the north--south density difference to which $q_N$ is proportional, and can completely collapse the MOC, differently from $E_s$, as discussed in the previous section.
The freshwater flux needed to collapse the MOC is about $0.4\Sv$, a value comparable to that used in several ``hosing'' experiments with models of various complexity (e.~g.~\cite{hawkins_bistability_2011}) and much smaller than the symmetric freshwater flux needed to collapse the J07 model.
Even if a similar freshwater flux $E_a$ is used as a ``hosing'' in the model of J07, the sensitivity in this model is still relatively weak.

In Fig.~\ref{fig:DqSqN_Ea}, the solution is continued to negative values of $q_S$, down to the point where the downwelling stops in the north.
This implies that, at $q_S=0$, Eqs.~\eqref{eq:box-modela}, \eqref{eq:box-modelb}, \eqref{eq:box-modeld} and~\eqref{eq:box-modele} are changed in order to account for the reversal of the advection direction at the boundaries of the boxes s and ts (see Sec.~\ref{sec:solution}).
The continuation to negative $q_S$, down to the point $q_N=0$, is performed for discussing the relevance of $M_{ov}$ as a stability indicator (Sec.~\ref{sec:Mov}).

At the point where $q_N$ reaches zero, an increase of the freshwater flux $E_a$ maintains the collapsed state with $q_N=0$ 
and a weakly negative $q_S$ compensating the upwelling in the Atlantic, as an increase of $E_a$ decreases $\rho_n$ with 
respect to $\rho_{ts}$. On the other hand, if starting from the point where $q_N$ reaches zero the freshwater flux is decreased 
further, the reversed solution can no longer exist, as a decrease of $E_a$ determines an increase of $\rho_n$ with respect to 
$\rho_{ts}$, which must be associated at the steady state with downwelling in the north. 
We thus conclude that the point defined by $q_N(E_a)=0$ is the second saddle node bifurcation $L_2$ of the bifurcation diagram of the MOC strength of the box model (Fig.~\ref{fig:BifurcationSketch}).

\subsubsection{Sensitivity of MOC stability to changes in the gyre circulation}
\label{sec:rSrN}

We now turn to the assessment of the sensitivity of the collapse position, the saddle node bifurcation $L_1$, to changes in the gyre circulation.
The position of the saddle node bifurcations of Fig.~\ref{fig:DqSqN_Ea} is tracked while $r_S$ or $r_N$ are changed.
This enables to compute the critical value of $E_a$ in a range of values of $r_S$ (Fig.~\ref{fig:Ea_rS}) and $r_N$ (Fig.~\ref{fig:Ea_rN}), for each value of $\eta$ used in the bifurcation diagram of Fig.~\ref{fig:DqSqN_Ea}.
The regime diagram obtained divides the parameter plane in two regions: below the critical curve an ``ON'' state of the MOC is possible, while above the line only a collapsed state is present.

Considering first the dependence of $L_1$ position on $r_S$ (top panel of Fig.~\ref{fig:Ea_rS}), we see that the asymmetric freshwater flux needed to collapse the MOC is reduced by an increase of the gyre strength in the south.
The decrease with $r_S$ of the freshwater forcing needed for collapsing the MOC is substantial, and leads almost to the disappearance of the ``ON'' state for the highest values of $r_S$ and lowest $\eta$.
Even if a gyre exchange of $50\Sv$ or more is unrealistic, the value of $\Maz$ associated with it (Fig.~\ref{fig:Ea_rS}, lower panel) is well within the range observed in GCMs \cite{drijfhout_stability_2010}.
A more effective way of changing $M_{az}$, i.~e.~changing the South Atlantic freshwater fluxes, may actually lead to the total disappearance of the ``ON'' state for realistic $E_a$ values \citep{cimatoribus_sensitivity_2011}.
\citet{longworth_ocean_2005}, on the other hand, did not observe this behaviour in their box model, finding that the effect of the gyres was always a stabilising one.
The difference in results is due to the fact that they only considered the impact of changes in both northern and southern gyres at the same time, in a much more symmetrical configuration (no periodic channel and ACC in the south, boxes of equal sizes in the north and south, and no box ts).

The sensitivity of the position of $L_1$ can be understood considering the lower panel of Fig.~\ref{fig:Ea_rS}, where the value of $E_a$ at $L_1$ is plotted versus $M_{az}$ instead of versus $r_S$.
The loss of stability of the ``ON'' state is connected to an increase of the amount of freshwater transported by the southern gyre into the Atlantic basin.
This reduces the north--south density difference by increasing the salinity of the box ts and slightly decreasing that of the box t; a smaller $E_a$ is then sufficient to bring $(\rho_n-\rho_{ts})$ to zero, stopping the sinking in the north.
We conclude that an increase of $M_{az}$, in this case obtained by increasing $r_S$, brings an effective negative density perturbation in the basin.
In terms of changes in the bifurcation diagram, we have a shift to the left of the bifurcation diagram and a slight decrease of the width of the ME region as $r_S$ increases (see Fig.~\ref{fig:Bif_rSrN}, top panel).

The impact of the northern gyre on the stability of the MOC (Fig.~\ref{fig:Ea_rN}) is opposite to that of the southern gyre.
A strengthening of the northern gyre increases the salinity in the northern box, increasing the MOC strength (not shown) and reducing the sensitivity of the MOC to freshwater perturbations.
This translates into two effects: (i) $L_1$ moves to higher values of $E_a$, i.~e.~a stronger freshwater flux from outside the Atlantic into the n box is necessary to collapse the MOC and (ii) the collapse of the MOC takes place at weaker MOC rates (Fig.~\ref{fig:Ea_rN}).
A strong gyre in the north brings salty waters in the n box independently of the state of the MOC, and thus reduces on one hand the effectiveness of a freshwater anomaly ($L_1$ moves to higher $E_a$), and on the other hand the importance of the salt--advection feedback ($L_1$ takes place at lower $q_S$ and $q_N$), as it reduces the relative importance of the salt transport by the MOC.
In connection with the latter point, also the width of the ME regime shrinks markedly (see Fig.~\ref{fig:Bif_rSrN}, lower panel).
This response to changes in $r_N$ is consistent with what was found by \citet{longworth_ocean_2005}.

\subsection{The role of $\mathbf{M_{ov}}$}
\label{sec:Mov}

In Fig.~\ref{fig:DqSqN_Ea}, the bifurcation diagram of $\Mov$ as a function of $E_a$ was shown, together with the MOC strength.
$\Mov$ monotonically decreases as the freshwater flux in the north is increased until $L_1$ is reached.
A decrease in $\Mov$ is thus a robust indication of a reducing distance from the limit point $L_1$, where the ``ON'' state is no longer stable.
It must be kept in mind that all the results presented here refer to steady state solutions.
Changes in $M_{ov}$ are thus differences between different steady states, and not trends in time.
The validity of the box--model approach, and that of $M_{ov}$, far from the steady state is unclear; for this reason we focus on the equilibrium solutions alone.

We can then consider the ME region, that is the part of the bifurcation diagram between $L_1$ and the end of the ``OFF'' branch, $L_2$.
In this region, a shut down state of the MOC is present besides the ``ON'' state under the same boundary conditions.
As discussed in the introduction, it has been suggested that the change in sign of $\Mov$ may mark the entrance into the ME regime, or in other words $\Mov$ may change sign at the value of $E_a$ for which $L_2$ is reached.
If $q_N=0$ is taken as $L_2$, we see that $M_{ov}$ is not a perfect indicator of the ME region, but it gives a good approximation of the ME region, at least for low values of $\eta$.

To analyse this issue in more detail, we study the difference between $\EaqN$, the value of $E_a$ for which $q_N=0$ and $\EaMov$, the value of $E_a$ for which $\Mov=0$ but $q_S,q_N>0$.
The first can be computed considering the system:
\begin{subequations}
  \begin{align}
    &q_N =0, \label{eq:EaqNa}\\
    &r_N (S_t-S_n)-(E_s+E_a) S_0 =0, \label{eq:EaqNb}\\
    &q_{Ek} S_s - q_e S_{ts} - q_S S_t + r_S (S_t - S_{ts}) = 0, \label{eq:EaqNc}\\
    &q_S S_s + q_e S_{ts} - q_{Ek} S_s - (E_s - E_a) S_0 = 0, \label{eq:EaqNd}\\
    &E_s-E_a =\Mov+\Maz, \label{eq:EaqNe}\\
    &q_U+q_S =0, \label{eq:EaqNf}
  \end{align}\label{eq:Eaq}
\end{subequations}
where the equations represent, in order, the condition of no downwelling in the north, the  salt budget of the northern box, the salt budget for the box ts, the salt budget for the box s, the freshwater budget of the Atlantic Ocean (the latter three for the case $q_S<0$, with $\Mov=q_S/S_0(S_d-S_t)$), and the volume budget for the pycnocline, all under the condition $q_N=0$.

From Eq.~\eqref{eq:EaqNf} the pycnocline depth at $L_2$ can be computed; choosing the positive solution of the second order equation we obtain:
\begin{equation}
  \DqN= \frac {1}{2} \frac{L_y}{L_{xA}} \frac{q_{Ek}}{A_{GM}} \left(1 + \sqrt{1 + 4 \frac{L_{xA}}{L_y} \frac{A_{GM} \kappa A}{q_{Ek}^2}} \right),
  \label{eq:DqN}
\end{equation}
where we write the value of $D$ at $L_2$ as $\DqN$, and the Ekman inflow into the pycnocline, $q_{Ek}=(\tau L_{xS})/(\rho_0 \left|f_S\right|)$, is left implicit.
Eliminating the salinities from Eqs.~\eqref{eq:EaqNa}--~\eqref{eq:EaqNe} and using Eq.~\eqref{eq:DqN}, an expression for $\EaqN$ can be obtained:
\begin{equation}
  \begin{split}
    \EaqN= E_s &\frac{r_N q_{Ek}-\qeqN \left(\qeqN -q_{Ek} +r_S \right)}{r_N q_{Ek} +\qeqN \left(\qeqN-q_{Ek}+r_S\right)} \\
    &+r_N \frac{\Delta T \alpha}{S_0 \beta}\frac{\qeqN \left(\qeqN-q_{Ek} +r_S\right)}{r_N q_{Ek} +\qeqN \left(\qeqN-q_{Ek}+r_S\right)},
    \label{eq:solEaqN}
  \end{split}
\end{equation}
where $\qeqN$ represents the value of the eddy flux at the point where $q_N=0$ and $\Delta T=T_{ts}-T_n$.
$\EaqN$ does not depend on $\eta$, as confirmed by the numerical results in Fig.~\ref{fig:DqSqN_Ea}.
It is interesting to compute the limit of $\EaqN$ for $r_S$ and $\kappa$ going to zero.
From Eq.~\eqref{eq:DqN}, we find that:
\[\lim_{\kappa \to 0} \DqN=\frac{L_y}{L_{xA}} \frac{q_{Ek}}{A_{GM}},\]
which simply states that if $\kappa$ goes to zero, $q_{Ek}=q_{e}$ if $q_N=0$, since also $q_S$ must be zero.
The non diffusive version of Eq.~\eqref{eq:solEaqN} is then easily obtained:
\begin{equation}
  \lim_{\kappa \to 0}\EaqN= E_s \frac{r_N - r_S}{r_N + r_S} + \frac{\Delta T \alpha}{S_0 \beta} \frac{r_N r_S}{r_N + r_S},
  \label{eq:solEaqNk0}
\end{equation}
from which the $r_S\to 0$ limit of $\EaqN$ can be computed:
\begin{equation*}
  \lim_{\kappa,\,r_S \to 0} \EaqN = E_s,
\end{equation*}
this shows that the ``OFF'' solution ends at the point where net evaporation over the Atlantic basin sums up to zero ($E_a=E_s$), if no vertical diffusion or gyre transport in the south are present: no ``OFF'' state is available for values of $E_a$ lower than $E_s$.
It is interesting to note that the presence of $r_N$ is irrelevant in this limit, i.~e.~the redistribution of freshwater within the Atlantic basin is irrelevant.
This limiting behaviour is the one discussed in \citet{rahmstorf_freshwater_1996}.

Similarly, the value of $E_a$ associated with $M_{ov}=0$, shorthanded $\EaMov$, can be obtained solving the system:
\begin{subequations}
  \begin{align}
    &S_{ts}=S_n \label{eq:EaMova}\\
    &q_N (S_t-S_n)+r_N (S_t-S_n)-(E_s+E_a) S_0=0 \label{eq:EaMovb} \\
    &q_{Ek} S_s -q_e S_{ts}-q_S S_{ts}+r_S (S_t-S_{ts}) =0 \label{eq:EaMovc}\\
    &q_S S_n+q_e S_{ts}-q_{Ek} S_s-(E_s-E_a) S_0 = 0 \label{eq:EaMovd}\\
    &q_S-q_N+q_U=0 \label{eq:EaMove}
  \end{align}
  \label{eq:EaMov}
\end{subequations}
where the first equation is the condition for $M_{ov}=0$ if $q_S>0$, the second is the salt budget for the northern box, the third is the salt budget for the ts box, the fourth is the salt budget for the s box (with $S_d=S_n$, as $q_S>0$) and the last equation is the volume budget of the pycnocline.
The algebra is in this case more tedious, but an expression for $\EaMov$ can be obtained similarly to what done for $\EaqN$, exploiting Mathematica software~\cite{wolfram_research_inc._mathematica_2010} (see Appendix~\ref{sec:appendix}).
The skill of $M_{ov}$ as an indicator of ME can then be measured as \[\Delta_E=\EaqN-\EaMov.\]
The analytical expression obtained for $\Delta_E$ and $\EaMov$ are very long, so that they are reported in the Appendix~\ref{sec:appendix}.
The expression of $\Delta_E$ in the limit $\kappa \to 0$ reads:
\begin{equation}
  \begin{aligned}
    \lim_{\kappa \to 0}\Delta_E &= \frac{r_S}{r_N+r_S} \left[ r_N \frac{\Delta T \alpha}{S_0 \beta} \right. \\
    &\left. -2  E_s \frac
      {q_{Ek}\, \Delta T \alpha\, \eta+ \frac{1}{2} \left(\frac{A_{GM} L_{xA}}{L_y}\right)^2 \left(1-\sqrt{1+4 \left(\frac{L_y}{A_{GM}L_{xA}}\right)^2 q_{Ek} \,\Delta T \alpha\, \eta}\right)}
      {\left(q_{Ek}+r_N+ r_S\right) \Delta T \alpha\, \eta+ \frac{1}{2} \left(\frac{A_{GM} L_{xA}}{L_y}\right)^2 \left(1-\sqrt{1+4 \left(\frac{L_y}{A_{GM}L_{xA}}\right)^2 q_{Ek}\, \Delta T \alpha\, \eta}\right)} \right],
    \label{eq:Diff0k0}
  \end{aligned}
\end{equation}
from which it trivially follows that $\Delta_E\to 0$ if also $r_S \to 0$.
We thus obtain the important result:
\begin{equation}
    \lim_{r_S,\,\kappa \to 0} \Delta_E =0,
  \label{eq:DeNoDiffNorS}
\end{equation}
recovering the result of the simpler model of~\citet{rahmstorf_freshwater_1996}.
Indeed, $\Mov$ is a perfect indicator of ME if the response of diffusive upwelling and southern gyre transport to MOC changes are neglected.
From Eq.~\eqref{eq:DeNoDiffNorS} we can expect $M_{ov}$ to be a good indicator of the ME regime as long as feedback mechanisms other than the salt--advection feedback, which is measured by $M_{ov}$, do not play a relevant role.
In particular, vertical diffusion and the freshwater transport by the gyre circulation (or by diffusion) at the southern border of the pycnocline must not provide an effective feedback mechanism to changes in the MOC, which in the simple framework of the box model is the limit $\kappa,r_S \to 0$.
In the limit of no vertical diffusion and no gyre in the south, other mechanisms are irrelevant to the skill of $\Mov$ as an indicator of ME.

\begin{figure}[pt]
  \begin{center}
    \includegraphics[width=0.7\textwidth]{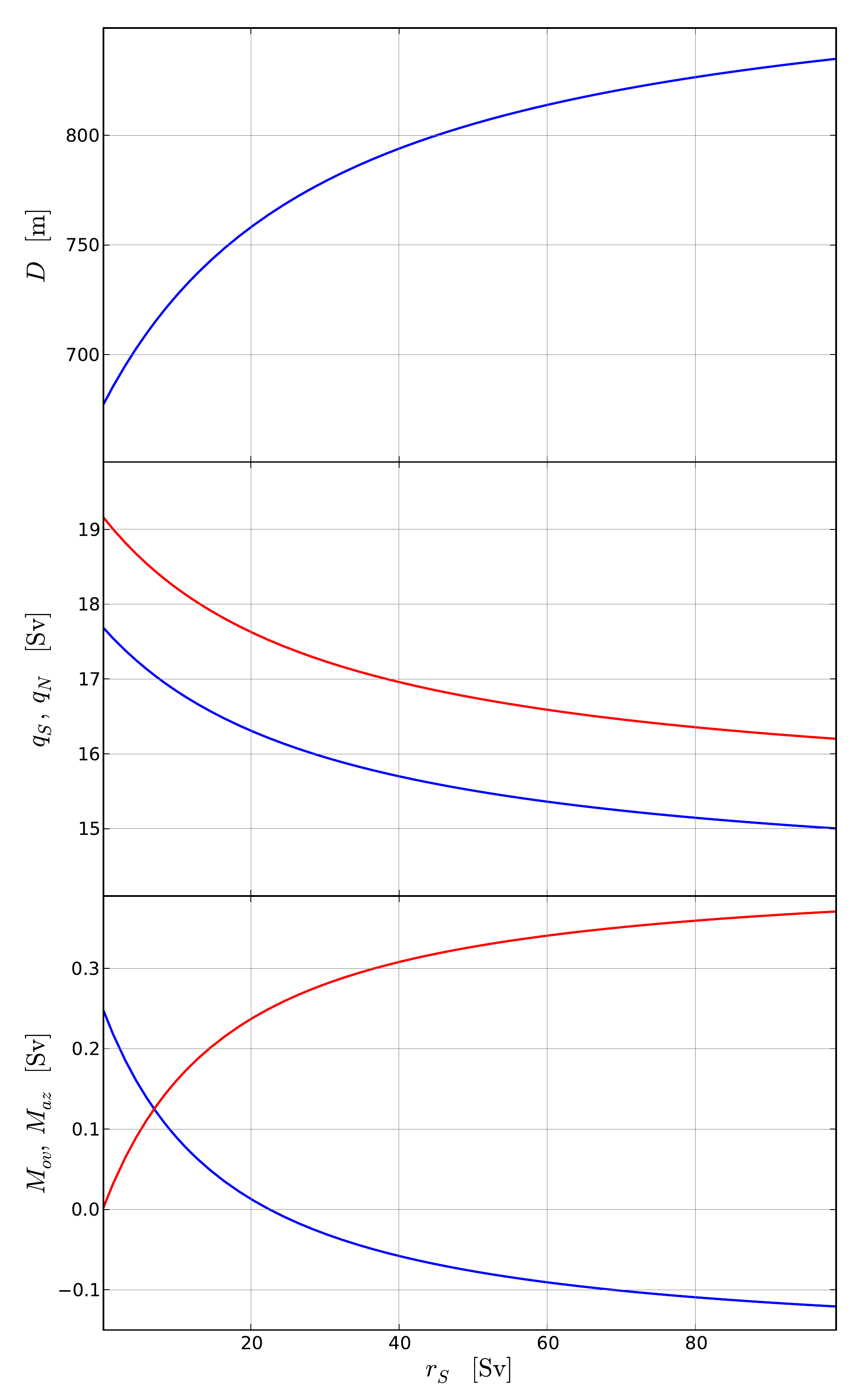}
  \end{center}
  \caption{Sensitivity of $D$ (top panel, blue), $q_S$ (central panel, blue), $q_N$ (central panel, red), $M_{ov}$ (lower panel, blue) and $M_{az}$ (lower panel, red) to the gyre exchange between the boxes ts and t ($r_S$).
    All other parameters are kept at the reference values of Tab.~\ref{tab:params}.}
  \label{fig:DqSqN_rS}
\end{figure}

\begin{figure}[ht]
  \begin{center}
    \includegraphics[width=0.7\textwidth]{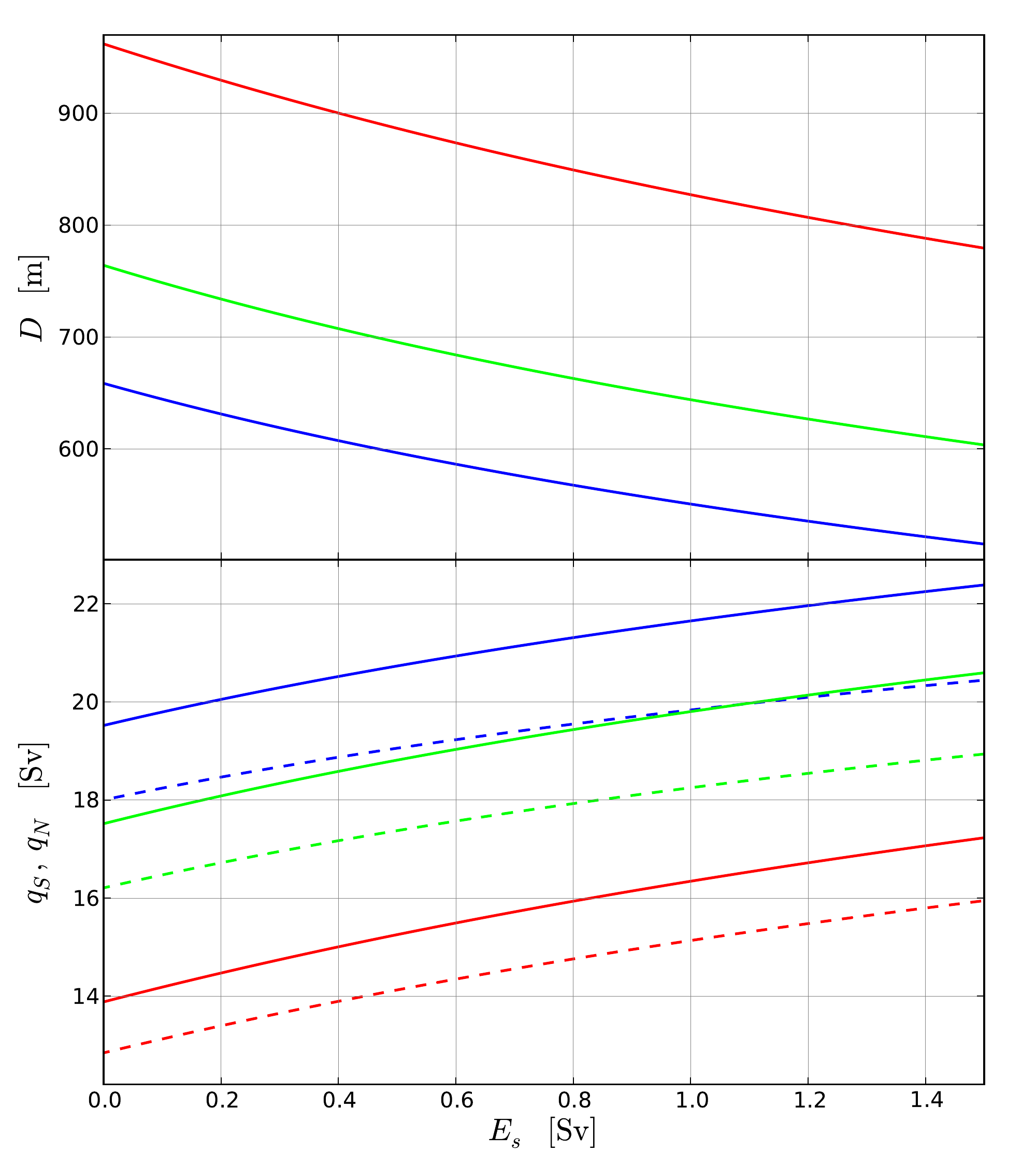}
  \end{center}
  \caption{Sensitivity of $D$ (top panel, blue), $q_S$ (central panel, dashed) and $q_N$ (central panel, full) to the symmetric freshwater flux $E_s$. The different colors (red, green and blue) refer to different $\eta$ values: $(1.5, 3.0, 4.5) \times 10^4\m\s^{-1}$ respectively.}
  \label{fig:DqSqN_Es}
\end{figure}

In Fig.~\ref{fig:Diff}, the dependency of $\Delta_E$ on $r_S$, $\kappa$ and $\eta$ are shown (from top to bottom).
The top panel shows the dependency of $\Delta_E$ on $r_S$ in the non--diffusive limit $\kappa \to 0$ and the centre panel that of $\Delta_E$ on $\kappa$ in the limit of $r_S\to 0$.
In the case of the southern gyre, $\Delta_E$ first grows towards negative values, reaching a minimum at about $5 \Sv$ to increase again and reach positive values for higher $r_S$ values.
An increase in $r_S$ has two effects: (i) it increases $\Maz$ and consequently decreases $\Mov$ (see Fig.~\ref{fig:DqSqN_rS}, bottom), making $\EaMov$ smaller, and (ii) it shifts $L_2$ to lower values of $E_a$ (see Fig.~\ref{fig:Bif_rSrN}, top).
The first change depends on how effective $r_S$ is in increasing $\Maz$, and from Fig.~\ref{fig:DqSqN_rS} it is clear that $\Maz$ increases quickly for low $r_S$, but then saturates for larger values of $r_S$, as the salinity difference between boxes t and ts decreases.
The value of $\EaqN$ decreases faster than $\EaMov$ at small $r_S$, but then saturates as well being almost constant above $r_S\approx 30\Sv$.
For this reason, at first the shift of $\EaqN$ dominates, but as $r_S$ increases the change in $\EaMov$ becomes more important, and thus $\Delta_E$ crosses zero a second time for $r_S\approx 25\Sv$, and is positive afterwards.
This second zero will move to higher values of $r_S$ as $\kappa$ increases, marking the position where the effects of $r_S$ and $\kappa$ compensate each other (not shown).
The values of the shift are always relatively small in magnitude ($<0.05\Sv$), as long as $r_S$ is not unrealistically large.

The dependency of $\Delta_E$ on $\kappa$ indicates that an increased vertical diffusion does not affect $\Mov$, but shifts $L_2$ towards more negative values of $E_a$, by stabilising the ``OFF'' state of the MOC and widening the ME region.
This translates into the shift of $\Delta_E$ towards negative values as $\kappa$ increases, as seen in Fig.~\ref{fig:Diff}.
This stabilising effect of vertical mixing within the Atlantic basin for the ``OFF'' state was already recognised by \citet{sijp_sensitivity_2006}.
In their study, they could not reach any permanently reversed state below a critical value of vertical diffusivity within the Atlantic basin.
This is not the case in our study.
This discrepancy is likely due to the use of perturbations too small to push the system to the ``OFF'' state with low diffusivity values in the work by \citet{sijp_sensitivity_2006}.
Considering the values of vertical diffusion used in most numerical models ($\mathcal{O}(10^{-4}\m^2\s^{-1})$), we may expect the skill of $\Mov$ in identifying the ME regime to be often low.
The fact that $M_{ov}$ has been instead successfully used to identify the ME regime in different numerical models (e.~g.~\cite{de_vries_atlantic_2005,hawkins_bistability_2011,huisman_indicator_2010}) is a sign that a compensation between the effects of vertical diffusion and horizontal advection (and/or diffusion) is taking place in numerical models.

When $\eta$ is increased, $\Delta_E$ goes from positive to negative values, if $r_S$, $\kappa$ or both are greater than zero (Fig.~\ref{fig:Diff}, bottom).
This is due to the proportionality between $M_{ov}$ and the MOC strength: increasing the hydraulic constant $\eta$ will increase $\Mov$, requiring a higher $E_a$ for bringing it to zero, while the $L_2$ position is independent of $\eta$  (Eq.~\eqref{eq:solEaqN}).
$\Delta_E$ is always smaller than $0.1\Sv$ if $\eta$ is taken in a range giving MOC rates not too far from reality ($\eta\approx 2-6\times 10^4\m\s^{-1}$).
If the gyre in the south and vertical diffusion are both set to zero, $\Delta_E$ is identically zero independently of $\eta$, as discussed above.

\section{Summary and conclusions}
\label{sec:Conclusions}

A box model for the MOC was developed, focusing on the effect of the exchanges between the Atlantic Ocean and Southern Oceans on the stability of the Atlantic MOC.
The model includes a shallow box of variable depth, providing a basic representation of the Atlantic pycnocline depth dynamics in the tropics and subtropics, similarly to what was done in~\cite{gnanadesikan_simple_1999} and J07.
The pycnocline depth is set by a balance between inflow from the Southern Ocean due to Ekman pumping, outflow due to baroclinic activity near the southern subpolar front, upwelling at low latitudes due to vertical diffusion and downwelling at the high northern latitudes.
Differently from previous studies, the scaling for the downwelling flux in the north depends on the density difference between the northern North Atlantic and the region above the pycnocline north of the ACC, which is represented by a separate box.
Furthermore, the transport (of salinity) between the ACC and the Atlantic is treated in a different way than in earlier studies: Ekman inflow and eddy outflow are considered separately for what concerns their associated salt transport.

Similarly to what obtained in J07 and \citet{gnanadesikan_simple_1999}, this configuration produces an inter--hemispheric MOC only if sufficiently strong wind stress is acting over the Southern Ocean. 
The MOC is, however, still buoyancy controlled, as changes in the north--south density difference, determined in the model only by changes in freshwater forcing or transport, can collapse the MOC.
The main advance with respect to previous works, is the ability to reproduce the sensitivity of the MOC to changes in the freshwater transport by the southern subtropical gyre in the Atlantic Ocean.
These results point to the fundamental importance of the region south of the tip of Africa and north of the ACC in determining the MOC stability.

In particular, the freshwater transport  at the latitude of the southern tip of Africa, by either the overturning or the azonal circulation, can change completely the response of the MOC to perturbations in the surface freshwater flux.
In this view, the net freshwater import into the Atlantic basin by the meridional overturning circulation, $M_{ov}$, is playing a fundamental role, being associated with the growth of perturbations due to the salt--advection feedback.
If the salt--advection feedback is the dominant feedback connected with a MOC collapse, the sign of $M_{ov}$ completely determines whether a permanent collapse is possible.
If other responses are important (gyre circulation, vertical diffusivity, atmospheric feedbacks or others not considered in this paper), the sign of $M_{ov}$ will not be a perfect indicator of multiple steady states anymore, its skill being dependent on the strength of the salt--advection feedback in comparison with other responses.
For a parameter set representative of the real ocean, we can expect $M_{ov}$ to be well below $0.1\Sv$ when entering the ME regime.
Considering the effect of noise and internal variability, this difference from zero is unlikely to be important.
On the other hand, if vertical diffusion is providing a strong feedback during a MOC collapse (in particular in numerical ocean models with high vertical diffusivity---both physical and numerical) the skill of $M_{ov}$ as indicator of ME may be reduced.
The observation that $M_{ov}$ actually is a good indicator in different numerical models \cite{de_vries_atlantic_2005,hawkins_bistability_2011,huisman_indicator_2010}, suggests that a compensating mechanism between different feedbacks is operating, in particular between the response due to vertical diffusivity and horizontal advection by the southern subtropical gyre.
Apart from its sign, a downward trend in $M_{ov}$ in response to a slowly varying external forcing \footnote{That is, changes slow enough to leave the system close to the steady state.} is a robust signal of an approach to the collapse point of the MOC, independently of the source of the freshwater perturbation (i.~e.~any term of the freshwater budget Eq.~\eqref{eq:MovMaz}).

The box model enables to speculate on the importance of the latitude at which $M_{ov}$ is computed: ``Why thirty degrees south?'' or, in other words, why should $M_{ov}$ be computed at the southern edge of the Atlantic Ocean?
One hypothesis is that the latitude of the southern tip of Africa marks the point north of which the MOC can be considered as a coherent flow with approximately constant water properties in numerical models.
Southward of this latitude and down to the northern end of the ACC, the interaction between the Ekman inflow from the Southern Ocean and the salt transport from the north by the subtropical gyre determines the density of the waters that flow into the Atlantic basin.
North of this point, salinity anomalies entering the Atlantic Ocean will eventually be advected to the northern downwelling regions where they will translate to a MOC strength anomaly (neglecting feedbacks other than the salt--advection feedback).
Whether these anomalies are subject to a positive feedback loop or a negative one depends on the sign of the freshwater transport at the southern entrance of the basin.
A second hypothesis, possibly related to the first one, concerns the presence of zonal boundaries north of the southern tip of Africa, which can support a zonal pressure gradient across the basin, controlling the MOC strength through planetary geostrophic balance~\cite{callies_simple_2012}.
Changes in this region may affect the entire basin, setting the boundary conditions for the MOC within the basin.

We have demonstrated that the box model is a useful diagnostic tool to understand results from GCMs, and that $\Mov$ can be a good indicator of the stability of the MOC, but one may ask if these results are relevant to interpret the results from high--resolution GCMs and, more importantly, observations from the real ocean.
In order for our conclusions to hold in those cases as well, the main assumption that must be satisfied is that the meridional overturning circulation should have a meridionally coherent response to freshwater perturbations on long time scales, at least within the Atlantic basin.
Finally, even if the role of meridional density gradients in driving the MOC is corroborated by various studies (see e.~g.~\cite{marshall_momentum_2011,sijp_key_2012}), the relevance of the scaling for the downwelling flux in an eddying ocean with complex geometry is still to be demonstrated.
While the downwelling in the subpolar gyre seems to be well described by the theory of \citet{spall_boundary_2004}, the connection between the downwelling in the subpolar gyre and the overturning circulation and its scaling law on the global scale is still unclear.

\begin{acknowledgments}
  A.A.C. acknowledges the Netherlands Organization for Scientific Research (NWO) for funding in the ALW program.
  The authors would like to thank three anonymous reviewers for their constructive, insightful and very useful comments.
\end{acknowledgments}

\appendix

\section{Derivation of $\mathbf{\EaMov}$ and $\mathbf{\Delta_E}$}
\label{sec:appendix}

To compute the $E_a$ value needed to bring $\Mov$ to zero with $q_S>0$, the system~\eqref{eq:EaMov} has to be solved.

In the non diffusive limit $\kappa \to 0$ Eq.~\eqref{eq:EaMove} reduces to $q_S-q_N=0$, a second order equation in $D$, which can be solved giving as positive solution:
\begin{equation*}
  \lim_{\kappa \to 0}\DMov=- \frac{1}{2} \frac{1}{\left(T_{ts}-T_n\right)\alpha\,\eta} \frac{A_{GM} L_{xA}}{L_y} \left[1-\sqrt{1+4\, q_{Ek}\left(\frac{L_y}{A_{GM} L_{xA}}\right)^2\left(T_{ts}-T_n\right)\alpha \,\eta} \right].
\end{equation*}
This result can be substituted into Eqs.~\eqref{eq:EaMova}--\eqref{eq:EaMovd}, and salinity can be eliminated giving:
\begin{equation}
  E_s \left[r_N - r_S + \left(T_{ts}-T_n\right) \left(\DMov\right)^2 \alpha\, \eta\right] =  E_a \left[r_N + r_S + \left(T_{ts} - T_n\right) \left(\DMov\right)^2 \alpha\, \eta\right].
  \label{eq:NoSalAPP}
\end{equation}
The latter is solved for $E_a$ giving $\EaMov$ in the case of $\kappa \to 0$:
\begin{equation}
  \lim_{\kappa \to 0}\EaMov=
  E_s\frac{r_N - r_S + \left(T_{ts}-T_n\right) \left(\DMov\right)^2 \alpha\, \eta}{r_N + r_S + \left(T_{ts}-T_n\right) \left(\DMov\right)^2 \alpha\, \eta}
  =E_s\frac{r_N - r_S +\qNMov}{r_N + r_S +\qNMov}.
  \label{eq:EaMov0NoDiff}
\end{equation}
\eqref{eq:EaMov0NoDiff} can be combined with Eq.~\eqref{eq:solEaqNk0} to give $\Delta_E$ in the non diffusive limit, Eq.~\eqref{eq:Diff0k0}.

The method outlined for the non diffusive limit can be used for the model including vertical diffusion as well.
Also in this case, the mathematics involved is very simple on conceptual grounds, but the large expressions obtained render the problem tedious with pencil and paper.
The final result is obtained with Mathematica software.
Eq.~\eqref{eq:EaMovd}, in this case a third order algebraic equation, can be solved for pycnocline depth, and the positive solution reads:
\begin{equation*}
  \begin{aligned}
    \DMov &= \frac{A_{GM} L_{xA}}{3 L_y \left(T_n-T_{ts}\right) \alpha \, \eta} \\
    -&\frac{\left(1+i \sqrt{3}\right) \left(A_{GM}^2 L_{xA}^2+3 L_y^2 q_{Ek} \left(T_{ts}-T_n\right) \alpha\,  \eta \right)}
    {\left[\begin{aligned}
          &3 \sqrt[3]{4}\, L_y \left(T_n-T_{ts}\right) \alpha\,  \eta  \left(2 A_{GM}^3 L_{xA}^3 + 9 A_{GM} L_{xA} L_y^2 q_{Ek} \left(T_{ts}-T_n\right) \alpha \, \eta \right. \\
          & -3 L_y^{3/2} \left(T_n-T_{ts}\right) \alpha\,  \eta  \left(9 A \,L_y^{3/2} \left(T_n-T_{ts}\right)^2 \alpha \, \eta \, \kappa - \sqrt{3}\left(T_n-T_{ts}\right)\left(A_{GM}^2 L_{xA}^2 L_y q_{Ek}^2 \right. \right.\\
          & +4 A\, A_{GM}^3 L_{xA}^3 \,\kappa - 18 A \,A_{GM} L_{xA} L_y^2 q_{Ek} \left(T_n-T_{ts}\right) \alpha \, \eta\,  \kappa \\
          & \left. \left. \left. - L_y^3 \left(T_n-T_{ts}\right) \alpha\,  \eta  \left(4 q_{Ek}^3 + 27 A^2 \left(T_n-T_{ts}\right) \alpha \, \eta \, \kappa^2 \right)\right)^{1/2}\right)\right)^{1/3}
        \end{aligned}\right]
    } \\
    & +\frac{1}{6 \sqrt[3]{2}L_y \left(T_n-T_{ts}\right) \alpha \,  \eta } \left(i\sqrt{3} -1\right) \left\{2 A_{GM}^3 L_{xA}^3+9 A_{GM} L_{xA} L_y^2 \,q_{Ek} \left(T_{ts}-T_n\right) \alpha\,  \eta \right. \\
    & -3 L_y^{3/2} \left(T_n-T_{ts}\right) \alpha \, \eta  \left[ 9 A\, L_y^{3/2} \left(T_n-T_{ts}\right) \alpha \, \eta \, \kappa +\sqrt{3} \left(-A_{GM}^2 L_{xA}^2 L_y \,q_{Ek}^2 \right. \right.\\
    & -4 A \, A_{GM}^3 L_{xA}^3 \,\kappa +18 A \, A_{GM} L_{xA} L_y^2 q_{Ek} \left(T_n-T_{ts}\right) \alpha \, \eta \, \kappa \\
    & \left. \left. \left. + L_y^3 \left(T_n-T_{ts}\right) \alpha \, \eta  \left(4 q_{Ek}^3 +27 A^2 \left(T_n-T_{ts}\right) \alpha \, \eta \, \kappa ^2\right)\right)^{1/2}\right]\right\}^{1/3}
  \end{aligned}
\end{equation*}

After eliminating salinity, Eqs.~\eqref{eq:EaMova}--\eqref{eq:EaMovd} reduce again to~\eqref{eq:NoSalAPP}, and $\EaMov$ is still given by Eq.~\eqref{eq:EaMov0NoDiff}, but with $\DMov$ and $\qNMov$ for the finite vertical diffusion case.

The difference between $\EaqN$ and $\EaMov$ then gives $\Delta_E$, which can be written as:
\begin{equation*}
  \begin{aligned}
    \Delta_E= 
    &\frac{2 E_s r_S}{r_N+r_S+ \left(T_{ts}-T_n\right)\left(\DMov\right)^2 \alpha \,\eta} \\
    &- \frac{A_{GM} \DqN L_{xA} \left[A_{GM} \DqN L_{xA} +L_y\left( r_S -q_{Ek}\right)\right]\left[r_N \left(T_n-T_{ts}\right)\alpha +2 E_s S_0 \beta \right]}{\left[\left(A_{GM} \DqN L_{xA}\right)^2+L_y^2 q_{Ek} r_N +A_{GM} \DqN L_{xA} L_y \left(r_S -q_{Ek}\right)\right] S_0 \beta} \\
    = & \frac{2 E_s r_S}{r_N +r_S+\qNMov} \\
    &- \frac{\qeqN\left(\qeqN +r_S -q_{Ek}\right)\left[r_N \left(T_n-T_{ts}\right) \alpha+2 E_s S_0 \beta\right]}{\left[\left(\qeqN\right)^2+ q_{Ek}r_N+ \qeqN\left(r_S-q_{Ek}\right)\right] S_0 \beta}.
  \end{aligned}
\end{equation*}

In the limit of no vertical diffusion, $\qeqN=q_{Ek}$ if $q_N=0$, and thus $\Delta_E$ reduces to:
\begin{equation*}
  \lim_{\kappa \to 0}\Delta_E= r_S \left[ \frac{2 E_s}{r_N + r_S + \qNMov} - \frac{2 E_s}{r_N +r_S} + \frac{r_N \left(T_{ts}-T_n\right) \alpha}{\left(r_N+r_S\right) S_0 \beta}\right],
\end{equation*}
which can be written in the form of Eq.~\eqref{eq:Diff0k0} when $\qNMov$ is written explicitly.


\clearpage

\begin{table}[tb]
  \begin{tabular}{ c | c || c | c || c | c}
    $V_{tot}$  & $3.0\times 10^{17}\m^3$        &
    $V_s$     & $9\times 10^{15}\m^3$        &
    $V_n$     & $3\times 10^{15}\m^3$          \\
    $A$       & $1.0\times 10^{14}\m^2$        &
    $\rho_0$  & $ 1027.5\kg\m^{-3}$            &
    $\alpha$  & $2 \times 10^{-4}\K^{-1}$       \\
    $\beta$   & $8 \times 10^{-4} \psu^{-1}$    &
    $T_0$     & $5 \C$                       &
    $S_0$     & $35 \psu$                     \\
    $L_{xS}$   & $3.0 \times 10^7\m$           &
    $L_{xA}$   & $1.0 \times 10^7\m$            &
    $L_y$     & $1.0 \times 10^6\m$            \\
    $f_S$     & $-10^{-4}\s^{-1}$              &
    $T_n$     & $5.0 \C$                     &
    $T_{ts}$   & $10.0 \C$                     \\
    $A_{GM}$   & $1700\m^2\s^{-1}$              &
    $E_s$     & $0.25 \times 10^6\m^3\s^{-1}$  &
    $E_a$     & $-0.1 \times 10^6\m^3\s^{-1}$  \\
    $r_N$     & $ 5  \times 10^6\m^3\s^{-1}$   &
    $r_S$     & $ 10  \times 10^6\m^3\s^{-1}$  &
    $\tau$    & $0.10\N\m^{-2}$                \\
    $\eta$    & $3.0 \times 10^4 \m \s^{-1}$  &
    $\kappa$  & $1 \times 10^{-5} \m^2\s^{-1}$ &
  \end{tabular}
  \caption{Reference values of the parameters used in Eqs.~\eqref{eq:PycnoclineNew} and~\eqref{eq:box-model}.\label{tab:params}}
\end{table}

\begin{table}[tb]
  \begin{tabular}{ c | c || c | c || c | c}
    $D$   & $696 \m$ &
    $S_n$ & $35.0 \psu$ &
    $S_d$ & $35.0 \psu$ \\
    $S_t$ & $35.2 \psu$ &
    $S_{ts}$ & $34.6 \psu$ &
    $S_s$ & $34.4 \psu$ \\
    $q_N$ & $18.8 \Sv$ &
    $q_S$ & $17.4 \Sv$ &
    $q_U$ & $1.4 \Sv$ \\
  \end{tabular}
  \caption{Steady solution for the reference configuration of the model (using the parameters in tab.~\ref{tab:params}).\label{tab:sol}}
\end{table}

\begin{figure}[pt]
  \begin{center}
    \includegraphics[width=0.7\textwidth]{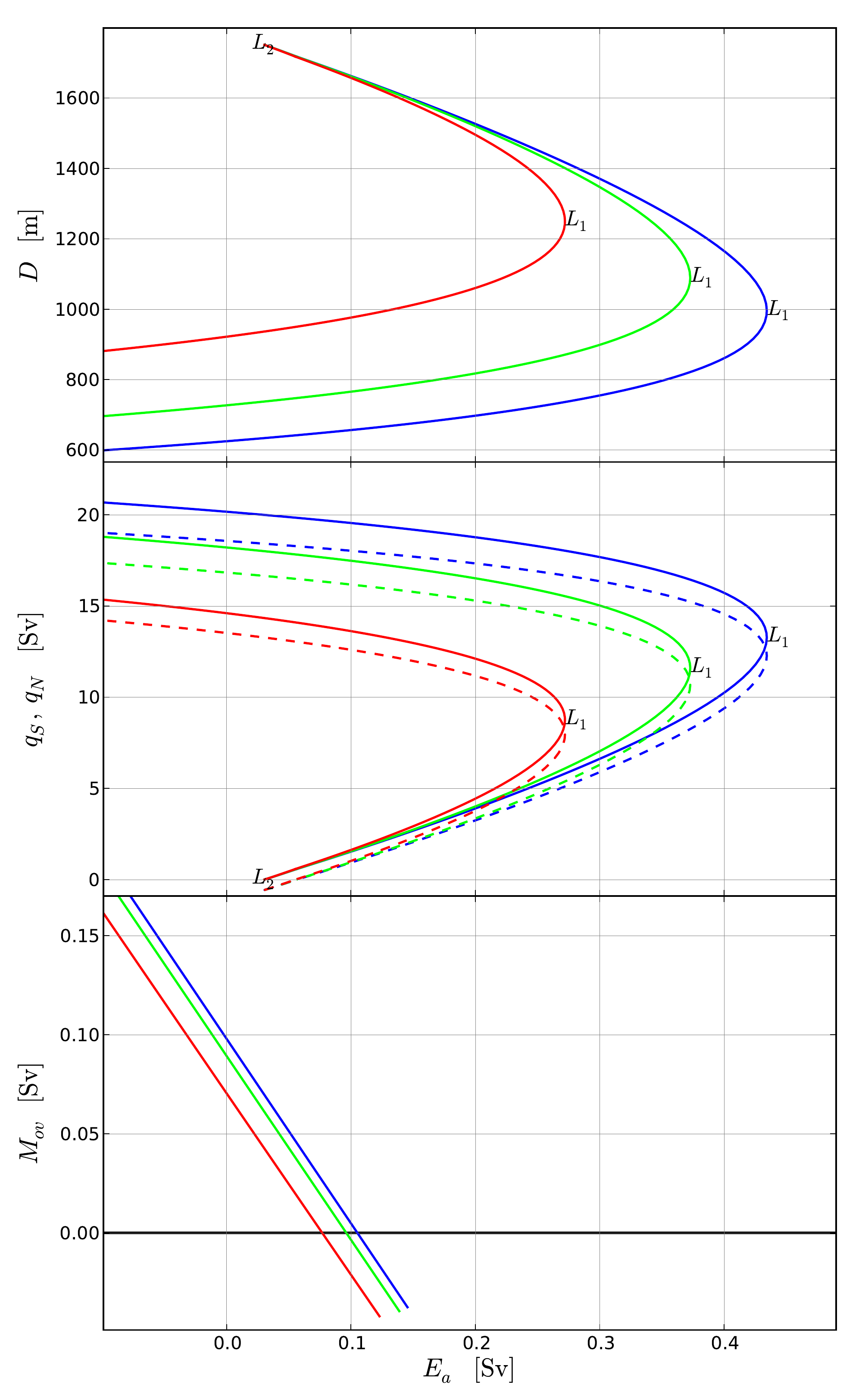}
  \end{center}
  \caption{Sensitivity of $D$ (top panel, blue), $q_S$ (central panel, dashed), $q_N$ (central panel, full) and $M_{ov}$ (lower panel) to the asymmetric freshwater flux $E_a$.
    The different colors (red, green and blue) refer to different $\eta$ values: $(1.5, 3.0, 4.0) \times 10^4\m\s^{-1}$ respectively.
    $L_1$ marks the position of the saddle node bifurcation ending the ``ON'' state of the MOC.
    After $L_1$, the solution is unstable, and for higher values of $E_a$ only a collapsed state of the MOC is possible.
    $L_{2}$ marks the position of the saddle node bifurcation ending the ``OFF'' state of the MOC.
    The plot of $M_{ov}$ stops before the saddle node bifurcations for clarity.
  }
  \label{fig:DqSqN_Ea}
\end{figure}

\begin{figure}[tb]
  \begin{center}
    \includegraphics[width=0.7\textwidth]{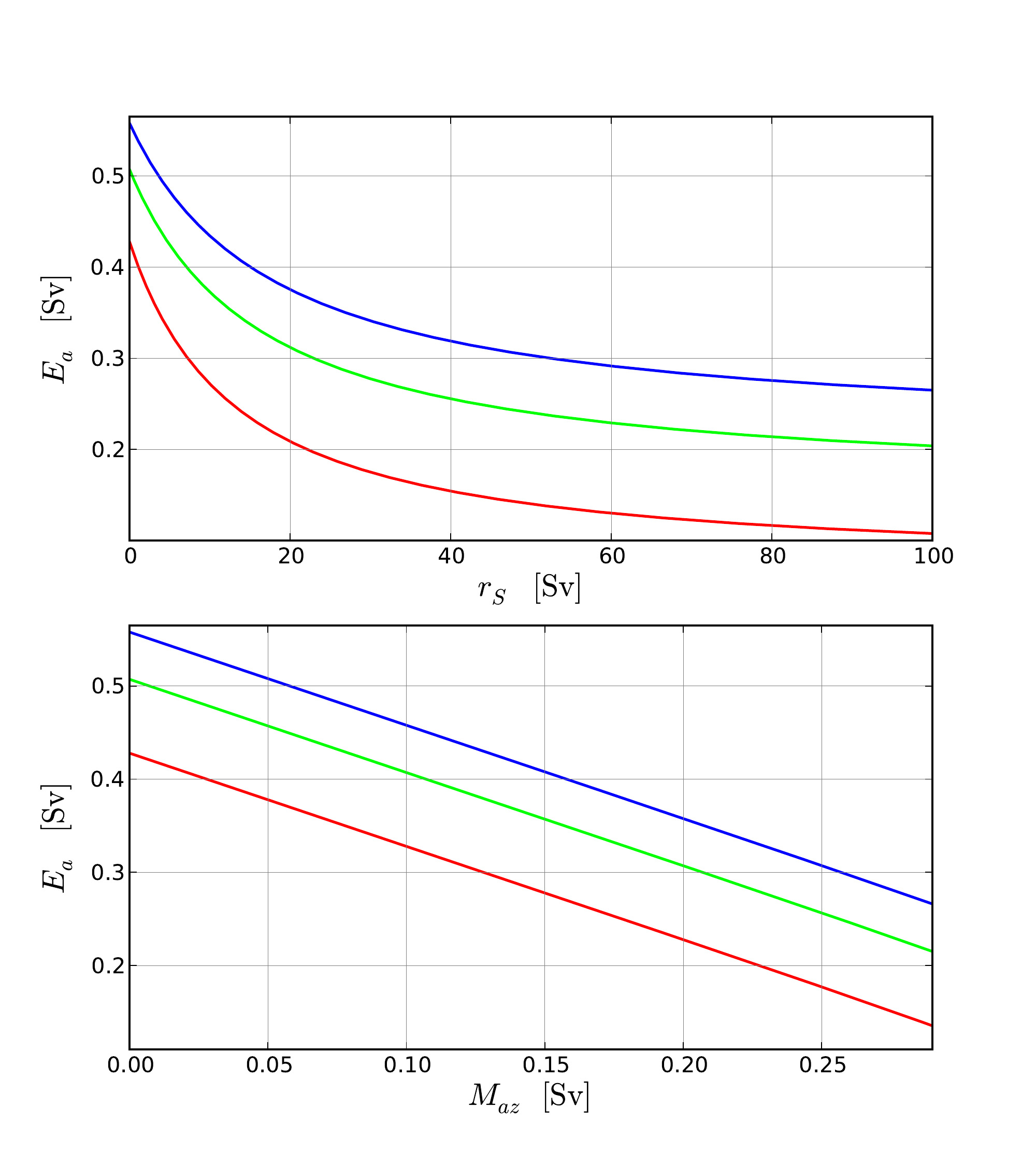}
  \end{center}
  \caption{Continuation of the limit point $L_1$, determining the collapse of the MOC, (see Fig.~\ref{fig:DqSqN_Ea}).
    A stable ``ON'' state exists only below the lines, marking the position of $L_1$.
    On the top panel, the continuation in $(E_a,r_S)$ is shown, while in the lower panel the same regime diagram is shown using $M_{az}$ instead of $r_S$ for plotting.
    The different colors (red, green and blue) refer to different $\eta$ values: $(1.5, 3.0, 4.5) \times 10^4\m\s^{-1}$ respectively.}
  \label{fig:Ea_rS}
\end{figure}

\begin{figure}[ht]
  \begin{center}
    \includegraphics[width=0.7\textwidth]{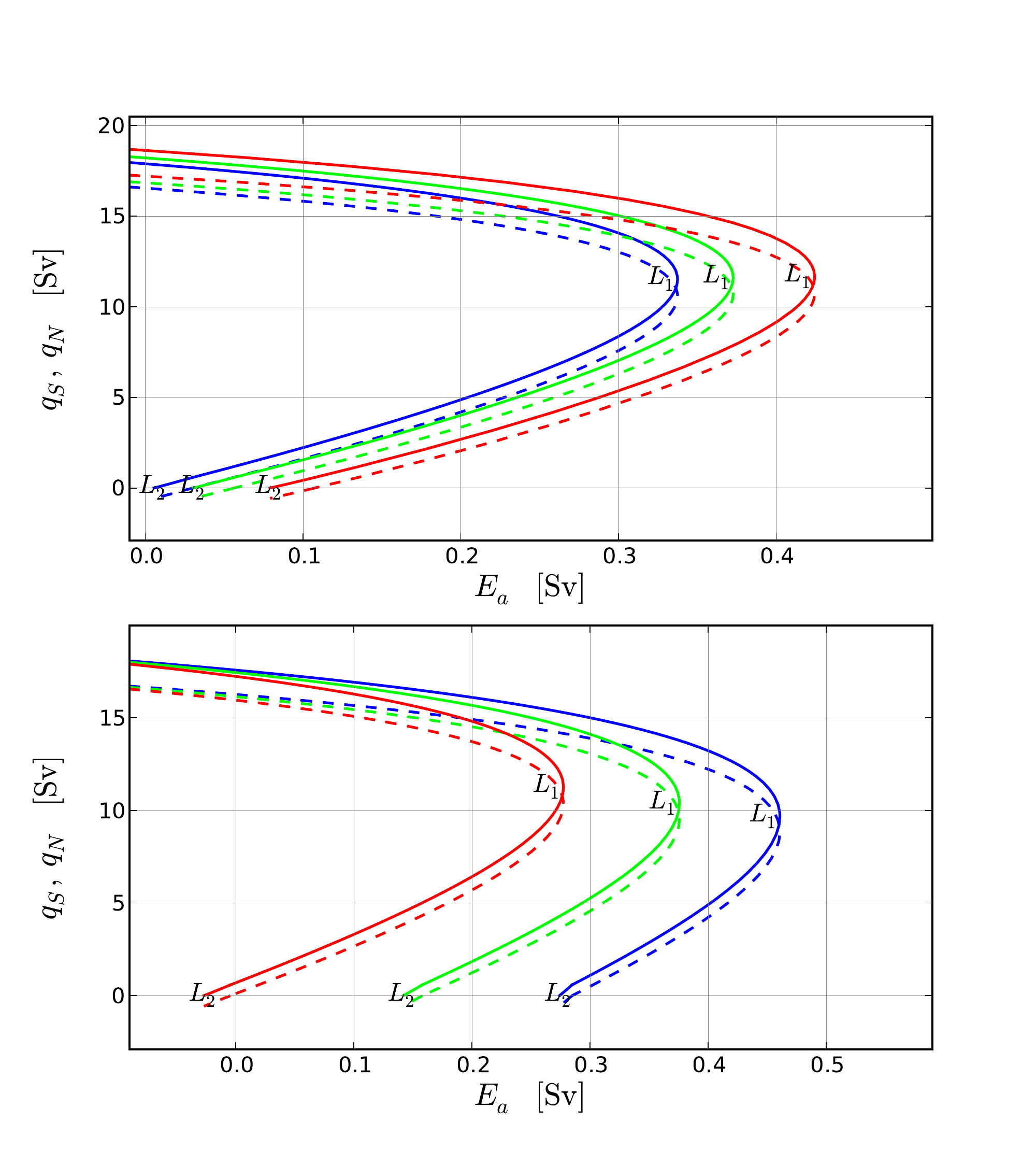}
  \end{center}
  \caption{Changes in the bifurcation diagram of the MOC as a function of $E_{a}$, as the strength of the gyres are changed; the full (dashed) line refers to $q_{N}$ ($q_{S}$).
    On the top panel, $r_{S}$ is changed, while in the lower panel $r_{N}$ is changed.
    The different colors (red, green and blue) refer to different $r_{S}$ or $r_{N}$ values: $5$, $10$, $15 \Sv$ respectively.
    For the case of $r_{N}$, the discontinuity in the MOC strength at the point where $q_{S}=0$ is due to the fact the when $q_{S}$ reverses the sensitivities of the MOC change abruptly.
    $L_{1}$ and $L_{2}$ mark the limit points ending the ``ON'' and ``OFF'' state respectively.
    \label{fig:Bif_rSrN}
  }
\end{figure}

\begin{figure}[ht]
  \begin{center}
    \includegraphics[width=0.7\textwidth]{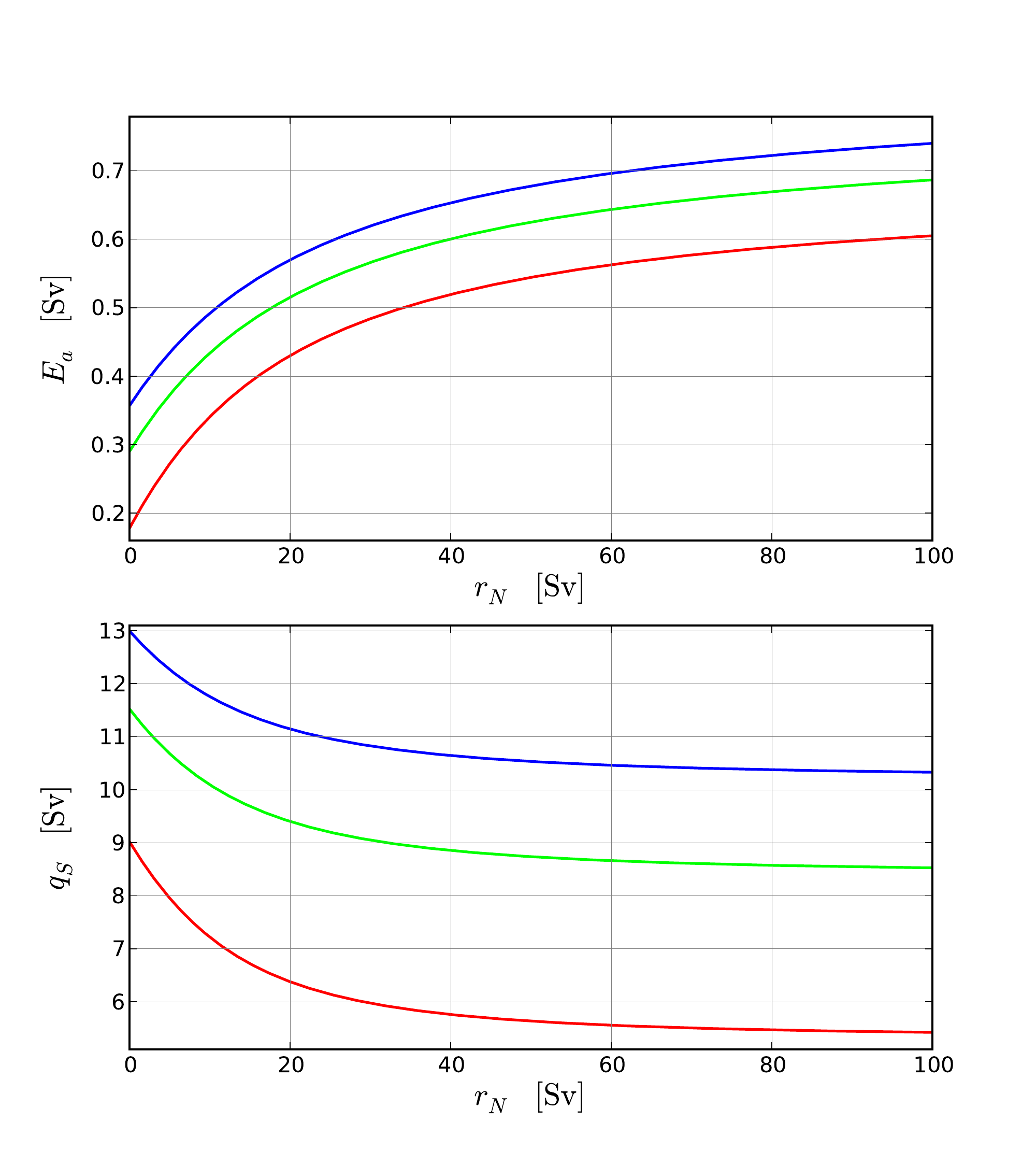}
  \end{center}
  \caption{Continuation of the limit point $L_1$ determining the collapse of the MOC (see Fig.~\ref{fig:DqSqN_Ea}).
    A stable ``ON'' state exists only below the lines, marking the position of $L_1$.
    On the top panel, the continuation in $(E_a,r_N)$ is shown, while in the lower panel the same regime diagram is shown considering $q_S$ instead of $E_a$, that is the strength of the flux in the south at the point of collapse.
    The different colors (red, green and blue) refer to different $\eta$ values: $(1.5, 3.0, 4.5) \times 10^4\m\s^{-1}$ respectively.}
  \label{fig:Ea_rN}
\end{figure}

\begin{figure}[pt]
  \begin{center}
    \includegraphics[width=0.7\textwidth]{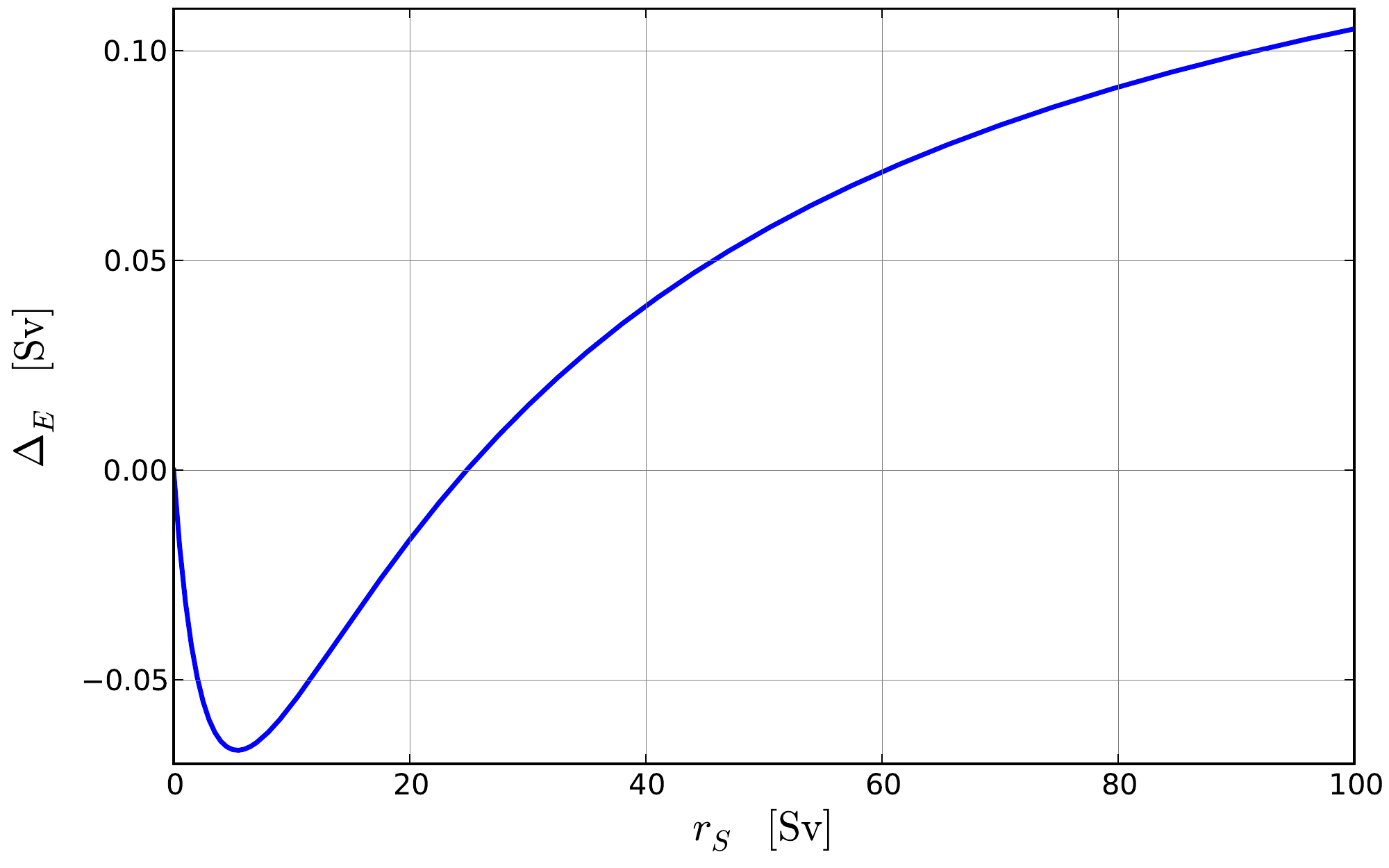}
    
    \includegraphics[width=0.7\textwidth]{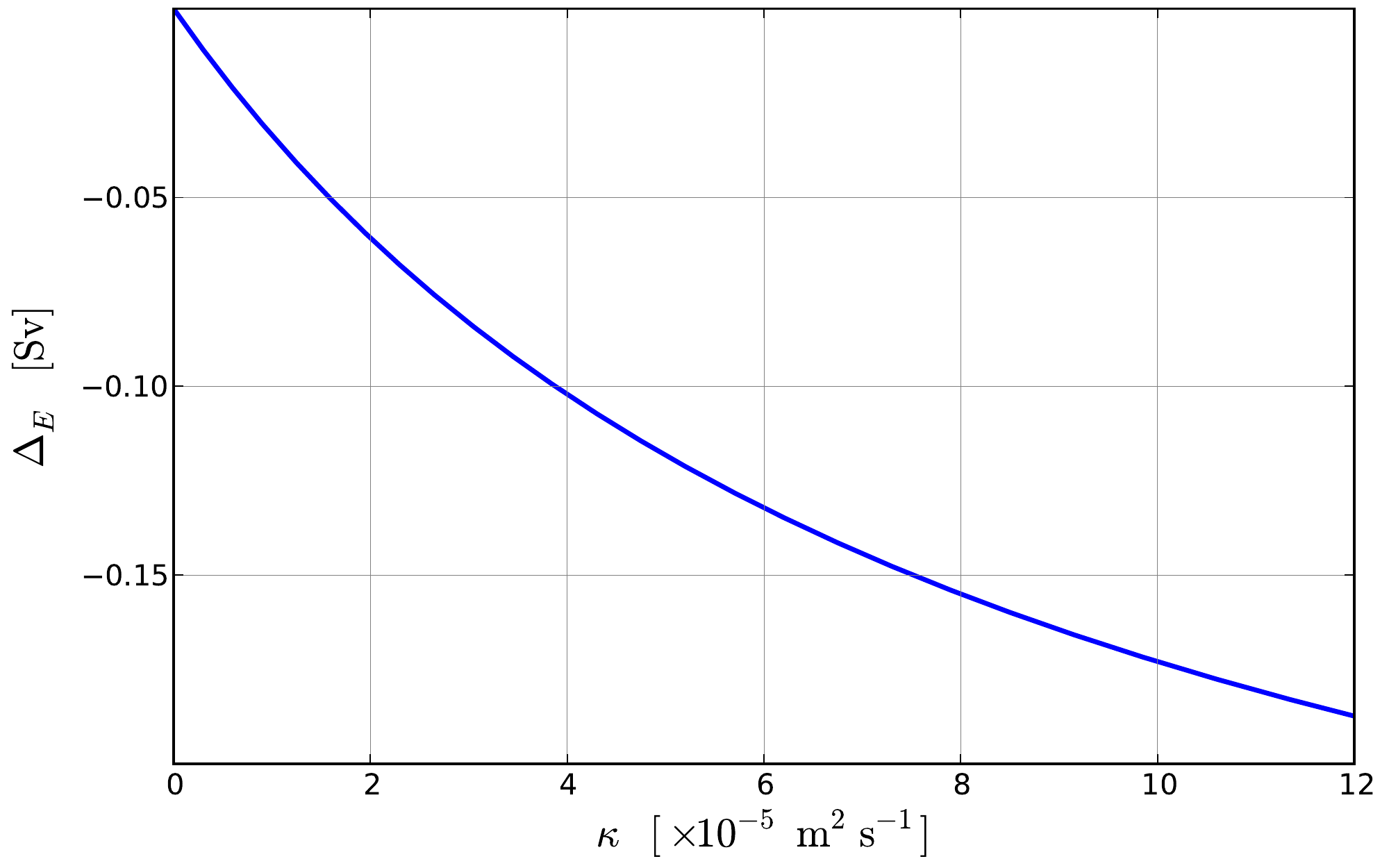}
    
    \includegraphics[width=0.7\textwidth]{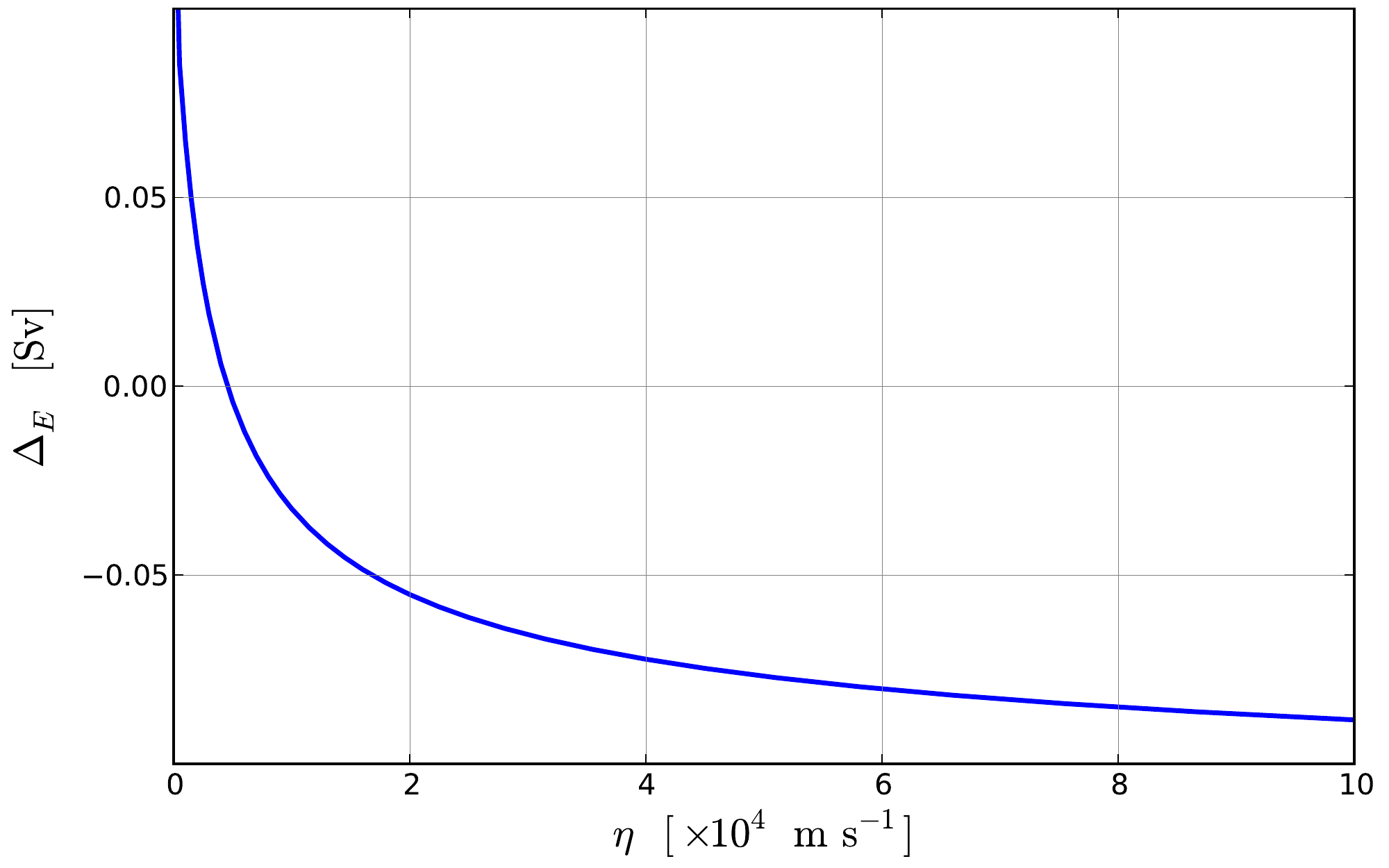}
  \end{center}
  \caption{Dependence of $\Delta_E$ on $r_S$ (top), $\kappa$ (centre) and $\eta$ (bottom).
    $\Delta_E$ is shown on the top (central) panel as a function $r_S$ ($\kappa$), keeping $\kappa=0\m^2\s^{-1}$ ($r_S=0\Sv$); all other parameters are kept at the reference value of Tab.~\ref{tab:params}.
    In the bottom panel, $\Delta_E$ is shown as a function of $\eta$ with all other parameters as in Tab.~\ref{tab:params}.
    \label{fig:Diff}}
\end{figure}

\end{document}